\documentclass[final,fleqn,twocolumn,times]{elsarticle}

\usepackage{multirow}
\usepackage{amsmath}
\usepackage{csquotes}
\usepackage{physics}
\usepackage{mathtools}
\usepackage{cases}

\newcommand\restr[2]{{
  \left.\kern-\nulldelimiterspace 
  #1 
  \right|_{#2} 
  }}

\usepackage{jmr_arxiv}
\usepackage{framed,multirow}

\usepackage{amssymb}
\usepackage{latexsym}

\def\vecr{{\mathbf{r}}}

\definecolor{newcolor}{rgb}{.8,.349,.1}
\usepackage[labelfont=bf,textfont=md]{caption}

\journal{Journal of Magnetic Resonance}

\begin{document}

\verso{N. Moutal \textit{et al}}

\begin{frontmatter}

\title{Localization regime in diffusion NMR: theory and experiments}

\author[1]{Nicolas \snm{Moutal}\corref{cor1}}
\ead{nicolas.moutal@polytechnique.edu}
\author[2,3]{Kerstin \snm{Demberg}}
\author[1]{Denis \snm{Grebenkov}}
\ead{denis.grebenkov@polytechnique.edu}
\author[2]{Tristan Anselm \snm{Kuder}}

\cortext[cor1]{Corresponding author:
Tel: +33 1 69 33 46 96
}

\address[1]{Laboratoire de Physique de la Mati{\`e}re Condens{\'e}e, Ecole Polytechnique, CNRS, IP Paris,
91128 Palaiseau, France}
\address[2]{Medical Physics in Radiology, German Cancer Research Center (DKFZ), Heidelberg, Germany}
\address[3]{Faculty of Physics and Astronomy, Heidelberg University, Heidelberg, Germany}


\begin{abstract}

In this work we investigate the emergence of the  localization regime for diffusion in various geometries: inside slabs, inside cylinders and around rods arranged on a square array.
At high gradients, the transverse magnetization is strongly attenuated in the bulk, whereas the macroscopic signal is formed by the remaining magnetization localized near boundaries of the sample. As a consequence, the signal is particularly sensitive to the microstructure.
 Our theoretical analysis relies on recent mathematical advances on the study of the Bloch-Torrey equation. Experiments were conducted with hyperpolarized xenon-129 gas in 3D-printed phantoms and show an excellent agreement with numerical simulations and theoretical predictions. Our mathematical arguments and experimental evidence indicate that the localization regime with a stretched-exponential decay of the macroscopic signal is a generic feature of diffusion NMR that can be observed at moderately high gradients in most NMR scanners.
\end{abstract}

\begin{keyword}
Localization regime; High gradients; PGSE
\end{keyword}

\end{frontmatter}
\section{Introduction}

Diffusion magnetic resonance imaging (dMRI) is a non-invasive technique which aims at unraveling the microstructural properties of a sample through the measurement of diffusion of spin-bearing nuclei (e.g., protons in water molecules) \cite{Callaghan1991a,Price2009a,Grebenkov2007a}. One may probe finer-scale details by increasing the magnetic field gradient $g$ or the $b$-value ($b\propto g^2$, see below), which characterizes the strength of the diffusion encoding. For example, Wedeen \textit{et al} employed very large $b$-values, up to $40~\rm{ms/\mu m^2}$, for the reconstruction of neuronal fiber pathways in the human brain \cite{Wedeen2012a}. In contrast, most theoretical approaches rely on small $b$ approximations. In fact, the assumption of small $b$-values allows one to expand the dMRI signal in powers of $b$ (cumulant expansion) and truncate it to its first terms \cite{Grebenkov2007a,Kiselev2017a,Froehlich2006a}. In particular, the first term of the cumulant expansion yields the classical Gaussian phase approximation for the signal $E=e^{-bD}$, where $D$ is the effective (or apparent) diffusion coefficient of spin-bearing particles in the sample that may differ from their intrinsic diffusion coefficient $D_0$. Note that the formula $E=e^{-bD_0}$ is exactly valid for free diffusion for any gradient profile and strength.

The validity of the cumulant expansion and related approximations was first questioned theoretically by Stoller \textit{et al.} \cite{Stoller1991a}, who showed that the cumulant expansion has a finite convergence radius and  that a new regime emerges at higher gradients. This so-called ``localization regime'' \cite{Swiet1994a} was observed experimentally by H{\"u}rlimann \textit{et al.} for water in slabs of width $160~\rm{\mu m}$ at gradients as small as $20~\rm{mT/m}$ and $b$-values of about $10~\rm{ms/\mu m^2}$ \cite{Huerlimann1995a}. It features a very strong deviation from the Gaussian phase approximation signal, with $-\ln(E) \propto (bD_0)^{1/3}$. Moreover, due to the high $b$-values the transverse magnetization far from the boundaries of the medium vanishes and only a thin layer of thickness $\ell_g=(\gamma g/D_0)^{-1/3}$ near the boundaries contributes to the measured signal (with $\gamma$ being the gyromagnetic ratio of the diffusing nuclei). Here, $\ell_g$ denotes the gradient length, which can be interpreted as the typical diffusion length traveled by nuclei during the time needed to reach uncorrelated phases. This feature makes the signal at high $b$-values particularly sensitive to the microstructure \cite{Grebenkov2018a}. The influence of the permeability of the boundaries was later studied in \cite{Grebenkov2014b,Grebenkov2017a} and it was shown that the localization of the magnetization near the boundaries enhances the dependence of the resulting signal on the permeability. The localization regime was shown to emerge when the gradient length $\ell_g$ is much smaller than (i) the diffusion length traveled by spins during the gradient sequence and (ii) the typical size of confining domains (e.g., width of the slab, in the experiment by H{\"u}rlimann \textit{et al}).


To our knowledge, the work by H{\"u}rlimann \textit{et al} \cite{Huerlimann1995a} on diffusion in a slab is the only undeniable experimental evidence of the localization regime. While deviations from the Gaussian approximation were abundantly observed in biological tissues and mineral samples \cite{Callaghan1991b,Niendorf1996a,Jensen2005a,Jacob2007a,Laun2011a,Kiselev2017a,Novikov2018b}, it is usually difficult to identify unambiguously the origin of these deviations. In other words, the observed deviations may be related to the localization regime, but may also originate from distinct populations of water with different effective diffusion coefficients, mixture of restricted and hindered diffusion, etc. The existence of the localization regime in more general domains and, in particular, in unbounded domains (e.g., extracellular diffusion) has not yet been addressed experimentally. In this paper, we present experimental data, numerical simulations and theoretical arguments showing the emergence of the localization regime in two complementary geometries: diffusion inside cylinders and diffusion outside an array of rods. We also treat a slab geometry as a reference case.

The paper is organized as follows. First, we present in Sec. \ref{section:theory} the theoretical ground that we rely on in order to interpret the experimental results.
Section \ref{section:material} describes the experimental setup and the numerical simulations. In Sec. \ref{section:results}, we show that our theoretical analysis and numerical computations are in excellent agreement with the experimental data. The characteristic stretched-exponential decay of the signal in the localization regime is observed at moderately high gradients and in various geometries, including unbounded diffusion with obstacles. We discuss the implications of these results and conclude the paper in Sec. \ref{section:conclusion}.

\section{Theory}
\label{section:theory}

The evolution of the transverse magnetization $M$ in the simplest case of a constant magnetic field gradient $g$ is described by the Bloch-Torrey equation \cite{Torrey1956a}:
\begin{equation}
\frac{\partial M}{\partial t} = (D_0 \nabla^2 - i\gamma g x)M\;,
\label{eq:Bloch_Torrey}
\end{equation}
where 
we chose the $x$-axis as the gradient direction.
The differential operator on the right-hand side of the equation, $D_0\nabla^2-i\gamma gx$, that we call the ``Bloch-Torrey operator'' in the following, is usually seen as a perturbation of the classical diffusion (or Laplace) operator $D_0 \nabla^2$. However, the $i\gamma g x$ term makes the whole operator non-Hermitian, which affects deeply its spectral properties \cite{Stoller1991a,Grebenkov2017a,Grebenkov2018b,Almog2018a}: its spectrum may become discrete or even empty in unbounded domains; the eigenvalues are complex; the eigenmodes are no longer orthogonal to each other (in the Hermitian sense) and they do not necessarily form a complete basis \cite{Helffer2013a,Moiseyev2011a}. These mathematical facts suggest that theoretical approaches where the gradient term is treated as a small perturbation may be insufficient to fully understand the signal formation.

In the following, we analyze the behavior of the transverse magnetization, hence the signal, at high gradients and relatively long encoding times, which are the conditions of emergence of the localization regime. This section summarizes and further extends the theoretical ground developed in \cite{Grebenkov2007a,Stoller1991a,Swiet1994a,Huerlimann1995a,Grebenkov2014b,Grebenkov2018a,Herberthson2017a}.
Throughout this paper, we focus on the Bloch-Torrey equation \eqref{eq:Bloch_Torrey} with the linear gradient, although similar features may be relevant to other spatial profiles of the magnetic field, e.g. the dipole field (see \cite{Ziener2012a,Ziener2019a} and references therein).

\subsection{Eigenmode decomposition}

Let us first consider a pulsed-gradient spin-echo sequence with gradient strength $g$, encoding pulse duration $\delta$, and no diffusion step between the two pulses (in other words, the time separation  $\Delta$ between two encoding pulses is equal to $\delta$). We first assume that the domain $\Omega$ in which diffusion takes place is bounded.

In bounded domains, the term $i\gamma g x$ is bounded and one can show that the complex-valued eigenmodes $v_n$ of the Bloch-Torrey operator form a complete basis, except at some exceptional values of $g$ (that we neglect in the following, see \cite{Grebenkov2018b}). 
We stress that these eigenmodes and the corresponding eigenvalues depend on the gradient $g$. Since the Bloch-Torrey operator is symmetric, the eigenmodes are ``rectanormal'' \cite{Stoller1991a} in the sense that, after appropriate normalization, they satisfy the following orthogonality relation (not to be confused with the Hermitian orthogonality relation, in which $v_m(\vecr)$ is replaced by its complex conjugate $v_m^*(\vecr)$):
\begin{equation}
\int_\Omega v_n(\vecr) v_m(\vecr) \,\mathrm{d}\vecr=\begin{cases}0 \quad \text{ if } n\neq m\\ 1 \quad \text{ if } n=m\end{cases}\;.
\label{eq:rectanormal}
\end{equation}
This allows one to project functions onto the \{$v_n\}_{n\geq 1}$ basis by using standard formulas.
Let us assume that the initial condition is a uniform magnetization after the exciting $90^\circ$ radio frequency (RF) pulse, i.e. $M(\vecr,t=0)=1/V$, where $V$ is the volume of the domain $\Omega$.
Hence we decompose the transverse magnetization as:
\begin{equation}
M(\vecr,t)=\frac{1}{\sqrt{V}}\sum_{n\geq 1} \mu_n v_n(\vecr)e^{-\lambda_n t}\;,
\label{eq:eigenmode_decomposition}
\end{equation}
where $\mu_n$ are complex coefficients given by
\begin{equation}
\mu_n = \frac{1}{\sqrt{V}}\int_\Omega v_n(\vecr) \,\mathrm{d}\vecr\;,
\end{equation}
and $\lambda_n$ are the corresponding eigenvalues: 
\begin{equation}
-(D_0\nabla^2 - i\gamma g x)v_n = \lambda_n v_n\;.
\end{equation}
We set the minus sign in front of the Bloch-Torrey operator to have eigenvalues with a positive real part, thus we sort $\lambda_n$ by increasing real part.

At the end of the first gradient pulse (i.e., at $t=\delta$), a refocusing $180^\circ$ RF pulse is applied, which is equivalent to applying a complex conjugation to the magnetization $M$:
\begin{align}
M(\vecr,\delta+0)&=\frac{1}{\sqrt{V}}\sum_{n\geq 1} \mu_n^* v_n^*(\vecr) e^{-\lambda_n^* \delta}\nonumber\\
&=\frac{1}{\sqrt{V}}\sum_{n\geq 1} \mu_n^* e^{-\lambda_n^* \delta} \sum_{m\geq 1} \beta_{n,m} v_m(\vecr)\;,
\end{align}
where we have decomposed each $v_n^*$ over the eigenmode basis $(v_m)_{m\geq 1}$, with complex coefficients $\beta_{n,m}$:
\begin{equation}
\beta_{n,m}=\int_\Omega v_n^*(\vecr)v_m(\vecr) \,\mathrm{d}\vecr\;.
\label{eq:beta}
\end{equation}
Finally, after the second gradient pulse, the transverse magnetization is equal to
\begin{equation}
M(\vecr, 2\delta)=\frac{1}{\sqrt{V}}\sum_{n,m \geq 1} \mu_n^*\beta_{n,m} v_m(\vecr) e^{-(\lambda_n^*+\lambda_m)\delta}\;,
\label{eq:m_complet}
\end{equation}
and the resulting signal can be computed from
\begin{equation}
E=\int_\Omega M(\vecr,2\delta) \, \mathrm{d}\vecr=\sum_{n,m \geq 1} \mu_n^*\beta_{n,m}\mu_m e^{-(\lambda_n^*+\lambda_m)\delta}\;.
\label{eq:E_complet}
\end{equation}
We stress again that the eigenvalues $\lambda_n$ and eigenmodes $v_n(\vecr)$ depend on the gradient $g$, and so do the coefficients $\mu_n$ and $\beta_{n,m}$.
\subsection{Localization regime and gradient length}
\label{section:theory_localization}
In the regime of long encoding times $\delta$, one can truncate the above decompositions to their first terms:
\begin{align}
M(\vecr,2\delta)&\approx \frac{1}{\sqrt{V}} \mu_1^*\beta_{1,1} v_1(\vecr) e^{-2 \mathrm{Re}(\lambda_1)\delta}\;,
\label{eq:m_truncated}
\\
E&\approx C_{1,1} e^{-2\mathrm{Re}(\lambda_1)\delta}\;,
\label{eq:E_truncated}
\end{align}
where the coefficient $C_{1,1}=\lvert \mu_1 \rvert^2 \beta_{1,1}$ typically decreases slowly with $g$ compared to the exponential factor.
This approximation is valid if the next-order terms are negligible. This may happen if the next eigenvalue is sufficiently far from $\lambda_1$, i.e., $\lvert e^{-\lambda_2\delta}\rvert \ll \lvert e^{-\lambda_1\delta}\rvert$, or if the weights $\mu_n^*\beta_{n,m}$ and $\mu_n^*\beta_{n,m}\mu_m$ (with $(n,m)\neq (1,1)$) are small enough. 
Note that the eigenvalues may come in complex conjugated pairs, in which case the above expressions \eqref{eq:m_truncated} and \eqref{eq:E_truncated} would involve four terms (two diagonal terms and two cross terms) with similar exponential decays $e^{-2\mathrm{Re}(\lambda_1)\delta}$. We show in Sec. \ref{section:theory_geometry} that these additional terms may produce oscillations in the signal on top of the overall decay \eqref{eq:E_truncated}. For clarity, we first focus on Eq. \eqref{eq:E_truncated} in this subsection.

In this regime, the behavior of the magnetization, and thus the signal, is then dictated by the dependence of $\mathrm{Re}(\lambda_1)$ on the gradient $g$. In \cite{Stoller1991a,Swiet1994a,Grebenkov2014b,Grebenkov2017a,Grebenkov2018b}, an asymptotic expansion for the $\lambda_n$ at high gradients was derived. In particular, the first term of the expansion of the first eigenvalue is universal:
\begin{equation}
2\mathrm{Re}(\lambda_1)=\lvert a'_1 \rvert \frac{D_0}{\ell_g^2}+O(\ell_g^{-3/2})\;, \qquad (g\to\infty)
\label{eq:lambda_1}
\end{equation}
where $\ell_g=(\gamma g/D_0)^{-1/3}$ is the gradient length and $a'_1\approx-1.02$ is the first zero of the derivative of the Airy function $\mathrm{Ai}(z)$. This formula is universal in the sense that it does not depend on the geometry of the domain. By combining Eqs. \eqref{eq:m_truncated} and \eqref{eq:lambda_1}, one gets the stretched-exponential behavior, which is a hallmark of the localization regime,
\begin{equation}
-\ln(E)\propto (\ell_\delta/\ell_g)^2 \propto (bD_0)^{1/3}\;, \quad (bD_0 \gg 1)
\label{eq:asymptotic}
\end{equation}
where $\ell_\delta=\sqrt{D_0\delta}$ is the diffusion length during the gradient pulse, and $b=2\gamma^2g^2\delta^3/3$ for this gradient profile.

In order to offer an interpretation for the gradient length $\ell_g$, let us consider several nuclei that start at the same position and diffuse freely (i.e., they do not encounter any boundary or obstacle) under the magnetic field gradient $g$. After a time $t$, the nuclei are typically spread over a distance of the order of $\sqrt{D_0t}$, and thus have accumulated a random phase difference of the order of $(\gamma g t) \sqrt{D_0 t}$. The nuclei have practically uncorrelated phases when this difference is comparable with $2\pi$, which yields a decorrelation time: $t_g=(\gamma^2 g^2 D_0)^{-1/3}$. Note that by time-reversal, $t_g$ can also be interpreted as the time beyond which several diffusing particles arriving at the same position have almost uncorrelated phases. The gradient length $\ell_g$ is then the typical distance traveled by these particles during time $t_g$: $\ell_g = \sqrt{D_0 t_g}$.

If the gradient length $\ell_g$ is much smaller than the typical size of the medium $\ell_s$ and the diffusion length $\ell_\delta$, then nuclei at any point not too close to a boundary have completely uncorrelated phases and the resulting magnetization is negligible. As the motion of the nuclei is restricted near the boundary, the remaining magnetization at the echo time is localized in a thin layer of thickness $\ell_g$ close to the boundary: this is the localization regime.
In regard to Eq. \eqref{eq:m_truncated}, the localization regime corresponds to the localization of the first eigenmodes of the Bloch-Torrey operator at high gradients over a width $\ell_g$ \cite{Grebenkov2014b,Grebenkov2018a,Grebenkov2018b}. This localization occurs where the gradient is orthogonal to the boundary, i.e. where the boundary prevents diffusion along the gradient direction the most.

\subsection{Dependence on the geometry}
\label{section:theory_geometry}

Although the first term of the high-gradient expansion \eqref{eq:lambda_1} of the eigenvalues of the Bloch-Torrey operator is universal, the next-order terms depend on the geometry of the domain, and more precisely on the curvature, permeability, and surface relaxivity of the boundary at the points where the localization occurs \cite{Grebenkov2014b,Grebenkov2017a,Grebenkov2018b}. Throughout the article, we discard permeability and surface relaxivity and consider only the effect of curvature. In two dimensions, if one denotes by $R_c$ the curvature radius at a localization point, then the eigenvalues associated to the eigenmodes localized near that point can be expanded as \cite{Grebenkov2018b}
\begin{align}
2\mathrm{Re}(\lambda_{k,l})&=\lvert a'_k \rvert \frac{D_0}{\ell_g^2}+(2l-1)\frac{D_0}{\lvert R_c \rvert^{1/2}\ell_g^{3/2}}\nonumber\\
&-\frac{\sqrt{3}D_0}{2\lvert a'_k \rvert R_c \ell_g}+O(\ell_g^{-1/2})\;,
\label{eq:lambda_1_next}
\end{align}
where $a'_k$ is the $k$-th zero of the  derivative of the Airy function, and $k,l$ are positive integers (the smallest eigenvalue is given by $k=l=1$). 
This formula applies to diffusion inside cylinders or outside parallel rods, which can be effectively reduced to diffusion in a two-dimensional geometry.
A generalization of Eq. \eqref{eq:lambda_1_next} to three-dimensional domains is expected to be
\begin{align}
\!\!\!\!\! 2\mathrm{Re}(\lambda_{k,l_1,l_2})&=\lvert a'_k \rvert \frac{D_0}{\ell_g^2}+\frac{D_0}{\ell_g^{3/2}}\left(\frac{2l_1-1}{\lvert R_{c,1} \rvert^{1/2}}+\frac{2l_2-1}{\lvert R_{c,2} \rvert^{1/2}}\right)\nonumber\\
&-\frac{\sqrt{3}D_0}{2\lvert a'_k \rvert\ell_g}\left(\frac{1}{R_{c,1}}+\frac{1}{R_{c,2}}\right)+O(\ell_g^{-1/2})\;,
\label{eq:lambda_1_next_3d}
\end{align}
where $R_{c,1}$ and $R_{c,2}$ are the principal radii of curvature of the boundary at the localization point and $k,l_1,l_2$ are positive integers. In the following we focus on the formula \eqref{eq:lambda_1_next} for two-dimensional geometries, which corresponds to our experimental setup (see Sec. \ref{section:material}).

In the third term of Eq. \eqref{eq:lambda_1_next} the sign of the curvature radius $R_c$ should be taken into account: positive for a concave boundary (e.g., interior of a cylinder) and negative for a convex boundary (e.g., exterior of a cylinder).
In the above expansion, the first term is related to the variation of the eigenmode along the gradient direction and the second term is related to the variation of the eigenmode in the orthogonal plane.

The real part of the spacing between the first and the next eigenvalue in Eq. \eqref{eq:lambda_1_next} is $2D_0/(\lvert R_c \rvert ^{1/2}\ell_g^{3/2})$ which means that although the localization regime emerges when $\ell_\delta \gg \ell_g$, the asymptotic decay \eqref{eq:E_truncated} is established at a longer time: $\ell_\delta \gg \lvert R_c \rvert^{1/4}\ell_g^{3/4}$ (assuming that $\ell_g < \lvert R_c \rvert$ in the localization regime, which is always the case for basic geometries).
In other words, in the intermediate regime $\ell_g \ll \ell_\delta \ll \lvert R_c \rvert^{1/4}\ell_g^{3/4}$, the transverse magnetization is localized at the boundary of the domain, but the truncated formulas \eqref{eq:m_truncated} and \eqref{eq:E_truncated} are not valid since one has to take into account other eigenmodes that are not suppressed by the exponential factor in Eq. \eqref{eq:m_complet}. These eigenmodes have typically the same profile in the gradient direction but different profiles in the orthogonal plane. Since these eigenmodes correspond to different eigenvalues, the signal is a superposition of different decaying functions (as in Eq. \eqref{eq:E_complet}) and cannot be reduced to a single stretched-exponential decay.

Another effect of the geometry is the overlapping of localization pockets. 
In fact, to each localization point (at which the gradient is perpendicular to the boundary) is associated a family of eigenmodes and eigenvalues described by Eq. \eqref{eq:lambda_1_next} at high $g$. As a consequence, the smallest eigenvalues from different families can contribute to the signal even at large $\delta$, whereas the overlap between corresponding eigenmodes may lead to oscillating patterns.

To illustrate this effect, let us consider the simple example of a slab of width $L$ (with the gradient direction being orthogonal to the slab). Note that this geometry is reduced to a one-dimensional interval of length $L$. In this setting, the eigenmodes of the Bloch-Torrey operator can be localized at either of two endpoints of the interval, but only if the gradient length $\ell_g$ is much smaller than $L$. In fact, as we stated in Sec. \ref{section:theory_localization}, $\ell_g$ represents approximately the width of the localized eigenmodes. For example, Stoller \textit{et al} computed that the first eigenmodes start to localize at each side of the slab when $\ell_g < 0.38 L$. Due to the left-right symmetry ($x\to -x$), the first two eigenmodes $v_1(x)$ and $v_2(x)$ satisfy the identity $v_2(x)=v_1^*(-x)$ and $\lambda_2=\lambda_1^*$, and are localized at each endpoint of the interval. Thus the first two eigenvalues have the same real part and, in the limit of large $\delta$, the magnetization may be represented by a superposition of $v_1$ and $v_2$:
\begin{align}
M(&x,2\delta)\approx \left(c_{1,1}+c_{2,1} e^{-2i \mathrm{Im}(\lambda_1)\delta} \right)e^{-2\mathrm{Re}(\lambda_1)\delta} v_1(x)\nonumber\\
&+ \left(c_{2,2} + c_{1,2}e^{2i \mathrm{Im}(\lambda_1)\delta}\right)e^{-2\mathrm{Re}(\lambda_1)\delta} v_2(x)\;,
\label{eq:m_overlapping}
\end{align}
with $c_{n,m}=\mu_n^*\beta_{n,m}/\sqrt{V}$,
and the signal is given by 
\begin{equation}
E \approx C e^{-2 \mathrm{Re}(\lambda_1)\delta}\;,
\label{eq:signal_localization}
\end{equation}
with
\begin{equation}
C=2(C_{1,1}+\mathrm{Re}(C_{1,2}))\;,
\label{eq:overlapping}
\end{equation}
where $C_{1,1}$ was previously introduced in Eq. \eqref{eq:E_truncated} and $C_{1,2}$ is an oscillating function of $g$ and $\delta$ given by
\begin{equation}
C_{1,2}=\mu_1^*\beta_{1,2}\mu_2 e^{2i\mathrm{Im}(\lambda_1)\delta}\;.
\label{eq:overlapping_B}
\end{equation}
The factor $2$ in \eqref{eq:overlapping} reflects the fact that two eigenmodes contribute to the signal.
For a slab, one can compute the following expansion for the imaginary part of the first eigenvalue \cite{Stoller1991a}:
\begin{equation}
2\mathrm{Im}(\lambda_1)=\pm \left(\gamma g L - \sqrt{3}\lvert a'_1 \rvert \frac{D_0}{\ell_g^2}\right)+O(\ell_g^{-3/2})\;,
\label{eq:lambda_im}
\end{equation}
where the $\pm$ sign is merely a matter of convention (given that $\lambda_2=\lambda_1^*$).
Note that the eigenmodes of the Bloch-Torrey operator in a slab are known explicitly and can be written in terms of Airy functions, see \cite{Stoller1991a,Grebenkov2014b,Grebenkov2017a}.

\begin{figure*}[t]
\centering
\includegraphics[width=0.49\linewidth]{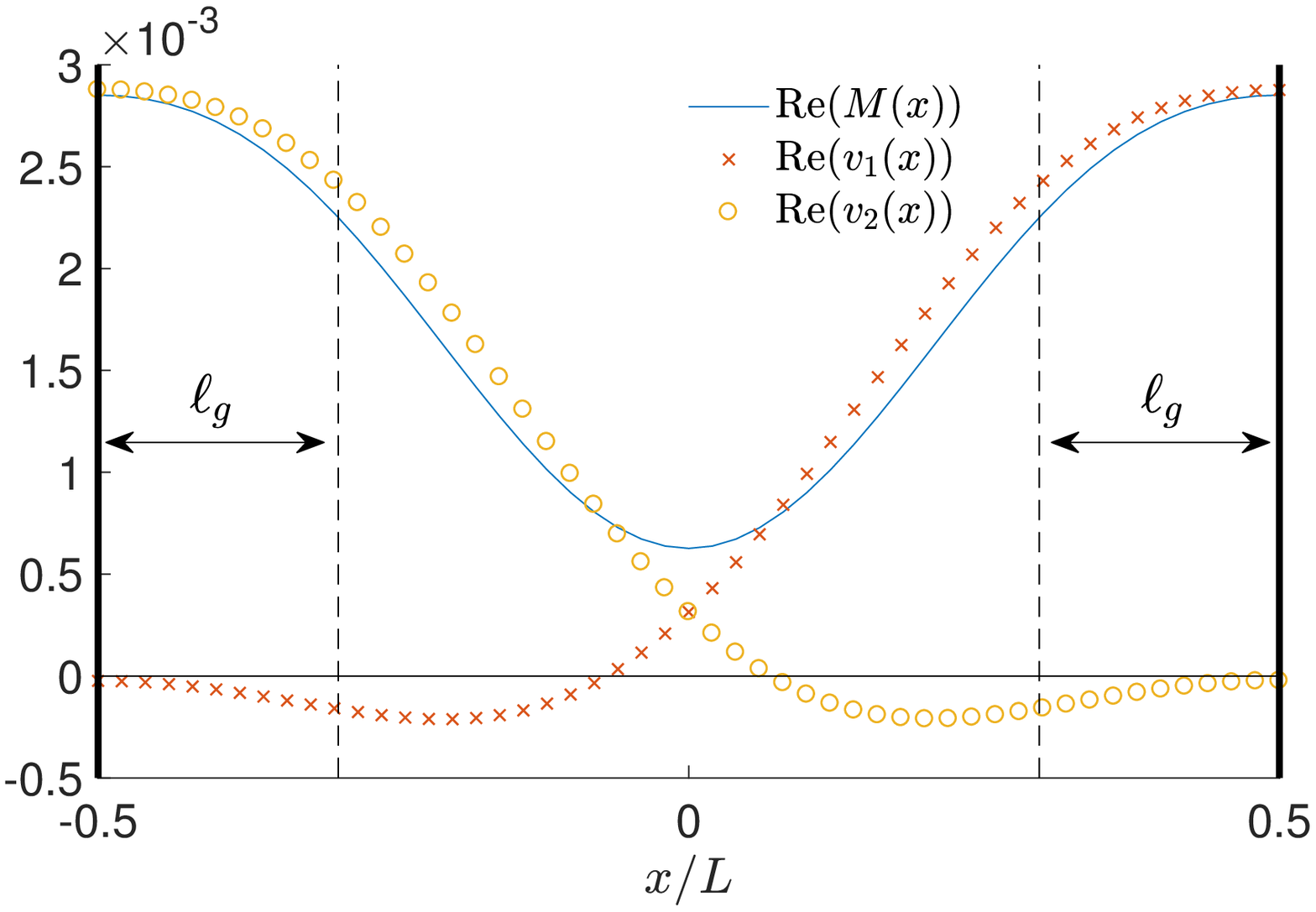}
\includegraphics[width=0.49\linewidth]{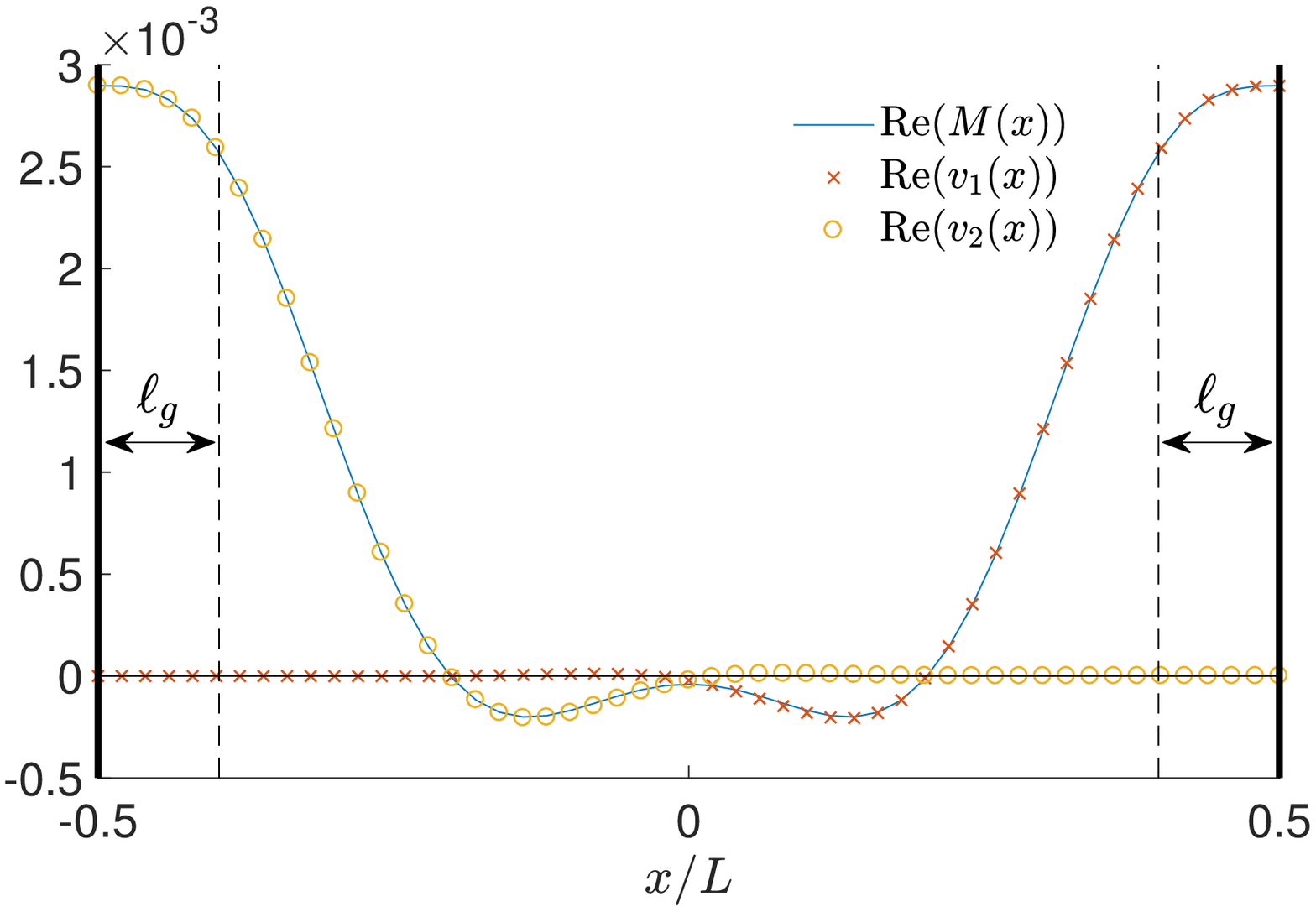}
\caption{The real part of the transverse magnetization $M(x)$ in a slab for two different gradient strengths, in the long $\delta$ regime, as well as the real part of the first two eigenmodes $v_1(x)$ and $v_2(x)$ weighted by the coefficients in front of $v_1(x)$ and $v_2(x)$ in Eq. \eqref{eq:m_overlapping}. We checked that Eq. \eqref{eq:m_overlapping}, i.e. the superposition of $v_1(x)$ and $v_2(x)$, reproduces perfectly the exact magnetization $M(x)$ in this regime. The computations were performed with a matrix formalism (described in Sec. \ref{section:material}). In both cases, one observes the localization of the magnetization near the endpoints of the interval. We chose a constant ratio $\ell_\delta/\ell_g=2.5$ for both figures, so that the amplitude of the magnetization is approximately the same (see Eq. \eqref{eq:asymptotic}).
(left) $\ell_g/L\approx 0.2$: one can see some overlapping between the two eigenmodes. (right) $\ell_g/L\approx 0.1$: there is almost no overlapping of the eigenmodes.}
\label{fig:overlapping}
\end{figure*}

The conclusion of this computation is that the cross-term $C_{1,2}$ produces oscillations in the signal on top of the asymptotic decay \eqref{eq:E_truncated}. The definition \eqref{eq:beta} of $\beta_{n,m}$ implies that this cross-term is linked to the overlapping of the modes $v_1$ and $v_2$. In turn, this overlapping depends on the ratio between the width $\ell_g$ of the modes, and their spacing $L$: the smaller the ratio $\ell_g/L$, the smaller the overlapping, and thus, the smaller the oscillating term. In the limit of well-separated modes, $\beta_{1,2}=0$ and one has $C=2C_{1,1}$. The effect of overlapping of eigenmodes is illustrated in Fig. \ref{fig:overlapping} where we plotted the transverse magnetization in a slab for two different gradient strengths: at low gradients the localization pockets overlap and at high gradients they are well-separated. Note that we plotted only the real part of the magnetization; the imaginary part is non-zero but it does not contribute to the signal since its integral over the slab is zero.
Although we illustrated this overlapping effect on the simple example of a slab, the conclusion (and the previous computations) may be generalized to any geometry, where $L$ would denote the spacing between two localized eigenmodes.

\subsection{The influence of $\Delta-\delta$}
\label{section:theory_diffusion}
Until now we have considered a PGSE sequence with $\Delta=\delta$, i.e. no diffusion step between two gradient pulses. In this section we consider a more general PGSE sequence and investigate the influence of the diffusion step on the transverse magnetization. The main parameter of this study is the duration of the diffusion step, $\Delta-\delta$. Mathematically, the effect of this diffusion step is to multiply the transverse magnetization just after the first gradient pulse by the operator $\mathcal{D}=\exp((\Delta-\delta)D_0\nabla^2)$. For simplicity we assume that the $180^\circ$ RF pulse occurs immediately after the first gradient pulse and not in the middle of the diffusion step (note that this assumption does not affect the final result). 
Then Eqs. \eqref{eq:m_complet}, \eqref{eq:E_complet}, and the consequent analysis remain applicable with the new definition for the coefficients $\beta_{n,m}$:
\begin{equation}
\beta_{n,m}=\int_\Omega (\mathcal{D}v_n^*)(\vecr) v_m(\vecr)\,\mathrm{d}\vecr\;.
\label{eq:beta_diff}
\end{equation}

The dependence on $\Delta-\delta$ is now hidden in the coefficients $\beta_{n,m}$. Note that if $\Delta=\delta$, then the operator $\mathcal{D}$ is the identity (since diffusion during zero time does not affect the magnetization) and we recover the previous definition \eqref{eq:beta} for $\beta_{n,m}$.
If we consider the regime of long $\delta$ where only the first eigenmode contributes, Eq. \eqref{eq:m_complet} is reduced to Eq. \eqref{eq:m_truncated} and  thus the diffusion step results merely in a modification of the global amplitude of the magnetization through the new definition \eqref{eq:beta_diff} of the coefficient $\beta_{1,1}$. In turn, the global amplitude of the signal in Eq. \eqref{eq:E_truncated} is also affected by this coefficient and decreases with increasing $\Delta-\delta$. 

The situation of two overlapping eigenmodes is more complex. The diffusion step also changes the values of the coefficients $\beta_{1,2}$, $\beta_{2,1}$, and hence the relative amplitudes of the oscillating term $C_{1,2}$ (see Eqs. \eqref{eq:overlapping} and \eqref{eq:overlapping_B}). Diffusion is expected to increase the width of the localized eigenmodes, thus increasing the overlapping between them and enhancing the oscillations in the signal. This effect should be stronger for already overlapping modes than for well separated ones, in other words, when the ratio $\ell_g/L$ is not very small, where $L$ denotes the distance between two localization pockets. The effect of the diffusion step on the localization regime in a one-dimensional setting was partly investigated in \cite{Grebenkov2014b}.

\subsection{Unbounded domains}

In the previous subsections, we assumed that diffusion takes place in a bounded domain $\Omega$, and this hypothesis ensured the validity of the decomposition of the transverse magnetization over the Bloch-Torrey eigenmodes basis \eqref{eq:eigenmode_decomposition}. However, this assumption excludes some physically relevant cases, such as extracellular diffusion, which are described by unbounded domains. In this section, we discuss few known mathematical properties and the applicability of the previous results to such domains.

The most striking property of the Bloch-Torrey operator is that, in most unbounded domains, its spectrum is discrete \cite{Almog2018a}, even though the spectrum of the Laplacian operator $\nabla^2$ is continuous. In other words, the $i\gamma g x$ term is responsible for the discreteness of the spectrum. However, there is no rigorous mathematical proof that the eigenmodes form a complete basis in such domains. We conjecture that this property still holds except for trivial cases (such as free space, which leads to an empty spectrum). In \cite{Grebenkov2018b}, a construction of approximate eigenmodes and eigenvalues was performed and Eqs. \eqref{eq:lambda_1_next} and \eqref{eq:lambda_im} are valid for unbounded domains as well.

Despite its practical importance, few theoretical studies addressed the validity of the Gaussian phase approximation in unbounded domains \cite{Grebenkov2018a}. In the following sections, we investigate experimentally the emergence of the localization regime in such a domain (array of rods).

\section{Material and Methods}
\label{section:material}

Experiments were performed using hyperpolarized xenon-129 gas ($\gamma\approx 74\cdot 10^6~\rm{s^{-1}T^{-1}}$) continuously flowing through phantoms containing different diffusion barrier geometries. Utilizing gas diffusion compared to water diffusion entails a several orders of magnitude larger diffusion coefficient which allows probing structures on the millimeter scale, which can easily be constructed with 3D-printers. Due to the weak signal of thermally polarized xenon gas, hyperpolarized gas with a considerably higher NMR signal was employed. Hyperpolarization was achieved by Rb/Xe-129 spin-exchange optical pumping (SEOP) \cite{Walker1997a,Oros2004a,Shah2000a}. For technical reasons \cite{Walker1997a}, a gas mixture (Air Liquide Deutschland GmbH, Düsseldorf, Germany) composed of xenon (0.95 Vol \%) and nitrogen (8.75 Vol\%) and helium-4 (rest) was used. The free diffusion coefficient of xenon in this gas mixture was measured to be $D_0 = (3.7 \pm 0.2) \cdot 10^{-5}~\rm{m^2s^{-1}}$ \cite{Kuder2013a}, which is one order of magnitude larger than for pure xenon gas \cite{Acosta2006a}. The gas was transferred at a small constant flow (approximately $150~\rm{mL/min}$) to the phantom positioned in an in-house built xenon coil in the isocenter of the magnet of a $1.5~\rm{T}$ clinical MR scanner (Magnetom Symphony, A Tim System, Siemens Healthcare, Erlangen, Germany) with a maximal employed gradient amplitude of $32 ~\rm{mT/m}$. The experimental set-up and hyperpolarization process are detailed in \cite{Kuder2013a,Demberg2017a}.

\begin{table}[t]
\centering
\begin{tabular}{|c|c|l|}
\hline
\# & Geometry & Dimensions \\
\hline
1 & slab & $L=3~\rm{mm}$ \\[0.5ex]
2 & cylinders & $2R=3.8~\rm{mm}$ \\[0.5ex]
3 & cylinders & $2R=2~\rm{mm}$ \\[0.8ex]
4 & array of rods & $17\times 10$ rods\\
& & $2R=3.2~\rm{mm}$; $L=4~\rm{mm}$ \\
 & & $b_V = 1.48~\rm{mm}$; $b_H = 1.35~\rm{mm}$\\[0.8ex]
5 & array of rods & $20\times  10$ rods\\
& & $2R=3.2~\rm{mm}$; $L=3.4~\rm{mm}$\\
 & & $b_V = 0.78~\rm{mm}$; $b_H = 1.08~\rm{mm}$\\
\hline
\end{tabular}
\caption{Characteristics of the phantoms (see Fig. \ref{fig:phantoms}).}
\label{table:phantoms}
\end{table}

\begin{figure}[t]
\centering
\includegraphics[width=0.99\linewidth]{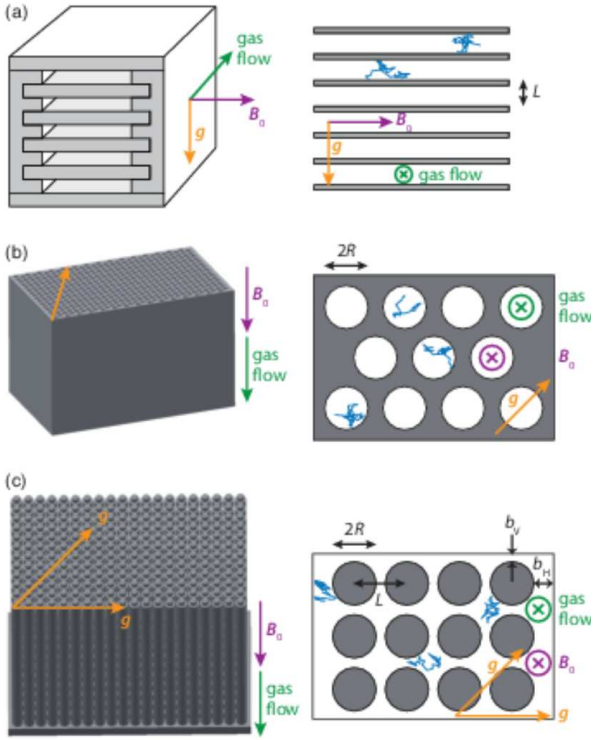}
\caption{Phantoms with corresponding magnified schematic depiction showing relevant length scales.
(a) Phantom $1$: Parallel plates. (b) Phantoms $2$ and $3$: Cylindrical tubes. (c) Phantoms $4$ and $5$: Cylindrical rods on a square grid.
}
\label{fig:phantoms}
\end{figure}

The phantoms used are illustrated in Fig. \ref{fig:phantoms} and described in Table \ref{table:phantoms}. Phantom $1$ contains parallel plates separated by a distance of $L=3~\rm{mm}$, built by the in-house workshop from PMMA. For the phantoms $2$ and $3$, two blocks containing cylindrical tubes (with two different diameters $2R$) in a hexagonal arrangement were 3D-printed. 
Here, the gas diffuses inside the cylinders. Since all cylinders are identical and isolated from each other, this setting is equivalent to a single cylinder of diameter $2R$. Phantoms $4$ and $5$ consist of cylindrical solid rods on a square grid attached to a base plate and a roof plate with holes for gas in and out flow, so that the gas diffuses outside the cylinders. The geometry is defined by the diameter of the rods ($2R$) and by the rod center-to-center distance ($L$).

Phantoms $2$-$5$ were printed with the PolyJet technology (Objet30 Pro, VeroClear as printing material, Stratasys, Ltd., Eden Prairie, MN, USA) and then inserted in a casing with a base area of $70.1~\rm{mm} \times 42.3~\rm{mm}$. Consequently, for phantoms $4$ and $5$, there are border regions between the array of rods and the enclosing walls in which diffusion also takes place. 
For all phantoms, surface relaxivity and permeation can be ignored for high-gradient experiments.

Phantom $1$ was positioned with the gas flow directed in the horizontal direction perpendicular to the main magnetic field and the gradient vector was pointing in the vertical direction in a sagittal slice of $50~\rm{mm}$ thickness orthogonal to the plates and to the gas flow direction, see Fig. \ref{fig:phantoms}. 
Phantoms $2$-$5$ were positioned with the gas flow directed parallel to the main magnetic field and the gradients were applied in the transversal plane of the scanner orthogonal to the gas flow direction (slice thickness $45~\rm{mm}$). For phantoms $4$ and $5$, measurements were taken with the gradient vector pointing along the left-right direction or the diagonal direction. 
In all cases, the gas flow did not influence the diffusion-weighted NMR signal.

\begin{figure}[t]
\centering
\includegraphics[width=0.99\linewidth]{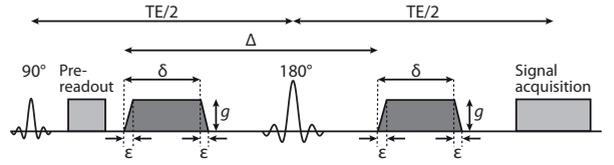}
\caption{PGSE sequence. The spin echo forms at time TE.}
\label{fig:PGSE}
\end{figure}

A pulsed-gradient spin-echo (PGSE) sequence was applied as depicted in Fig. \ref{fig:PGSE}. The durations $\delta$ of the trapezoidal gradient pulses were set to $6~\rm{ms}$ and include flat top time plus the ramp-up time of $\epsilon = 0.32~\rm{ms}$. The gradient separation time was $\Delta = 9.34~\rm{ms}$. The diffusion-weighted signal was sampled by  gradually increasing the gradient amplitude in 32 steps from 0 to $g_{\rm{max}} = 32~\rm{mT/m}$, recording the spin echo signal and acquiring up to $15$ averages. The time between two consecutive $90^\circ$ excitation pulses, i.e. the time between two measurements, was set to $18~\rm{s}$ to restore the polarization in the phantom via gas flow. To account for fluctuations in the polarization level the recorded spin echo was averaged and normalized to an additional signal pre-readout directly after the $90^\circ$ excitation pulse. To obtain the diffusion-induced signal attenuation, all points were normalized to the point acquired without diffusion weighting (i.e., at $g=0$). 

The numerical computation of the signal in a slab and in a cylinder is effectively reduced to that in an interval and a disk, respectively. For these simple shapes, the most efficient and accurate computation of the signal is realized with the matrix formalism \cite{Caprihan1996a,Callaghan1997a,Barzykin1998a,Grebenkov2007a}, in which the Bloch-Torrey equation is projected onto the basis of explicitly known Laplacian eigenfunctions to represent the signal via matrix products and exponentials (see \cite{Grebenkov2008b,Grebenkov2010a} for details). The matrix formalism was also used to compute the transverse magnetization in these two domains (see similar computations in \cite{Grebenkov2014b}). In turn, the numerical computation of the signals for phantoms $4$ and $5$ (arrays of rods) was performed differently. While the matrix formalism could in principle be applied, the need for a numerical computation of Laplacian eigenfunctions in such structures makes this approach less efficient. Thus, we performed Monte Carlo simulations including the borders around the rod arrays with $2.5\cdot 10^7$ random walkers and $1.6\cdot 10^5$ steps per random walk trajectory. In order to compute the eigenmodes of the Bloch-Torrey operator in the rods geometry, we used the PDE solver from Matlab (The MathWorks, Natick, MA USA) in a square array of $3\times 3$ rods with Dirichlet boundary conditions on the outer boundary and we kept only the eigenmodes that were localized on the central rod. Since the distance from the central rod to the outer boundary is about $L$, which was much larger than $\ell_g$ and $\ell_\delta$ in our simulations, the effect of the outer boundary is negligible, so that the computed eigenmodes are very close to the ones for the infinitely periodic array of rods.


%
%
%

\section{Results}
\label{section:results}

We present the experimental and numerical results for diffusion in three different geometries: inside slabs, inside cylinders, and outside arrays of rods. The localization regime in the slab geometry was already investigated experimentally by H{\"u}rlimann \textit{et al} \cite{Huerlimann1995a}, whereas only a few theoretical studies were devoted to the cylinder geometry \cite{Swiet1994a,Grebenkov2018a,Grebenkov2018b}. The signal for an array of rods was previously computed numerically in \cite{Grebenkov2018a} using a finite-element method \cite{Nguyen2014a}.

For our particular gradient sequence and the parameters of the xenon gas mixture, one can compute $\ell_\delta=0.5~\rm{mm}$ and $\ell_g$ decreases from $0.8~\rm{mm}$ to $0.25~\rm{mm}$ for $g$ ranging from $1~\rm{mT/m}$ to $32~\rm{mT/m}$. So, by increasing the gradient, $\ell_g$ crosses $\ell_\delta$ and the localization regime emerges.

\subsection{Slab geometry}

We choose the axes such that the slab is orthogonal to the $z$-axis. Note that this convention is different from the one that we adopted in Sec. \ref{section:theory}, where the gradient was directed along the x-axis.
One can then decompose the diffusive motion independently along the three axes $x,y,z$ and get
\begin{equation}
E=\exp(-bD_0 \sin^2\!\theta)\,E_{\rm slab}(L,g \cos\theta)\;,
\label{eq:angle_slab}
\end{equation}
where $\theta$ is the angle between the gradient and the $z$-axis and $E_{\rm slab}(L,g)$ is the signal from an interval of length $L$. 
Here, the first factor is the signal attenuation due to diffusion in the lateral plane $xy$, which is almost free as outer boundaries are separated by distances that greatly exceed both $\ell_g$ and $\ell_\delta$ (about $40~\rm{mm}$).
One gets the slowest decay by ensuring that the gradient is orthogonal to the slab, i.e. $\theta=0$, which was chosen in the experiments and numerical simulations. This is expected since in this situation the boundary restricts diffusion along the gradient direction the most.

\begin{figure}[t]
\centering
\includegraphics[width=0.99\linewidth]{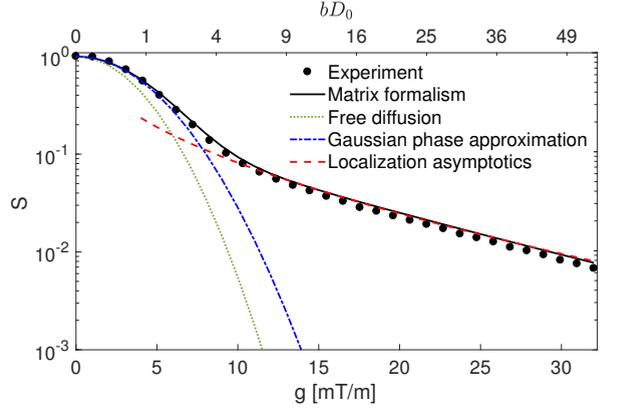}
\caption{Signal attenuation for phantom $1$ (a slab of width $3~\rm{mm}$). Experimental results are shown by full circles and matrix formalism computation by a solid line. The signal for free diffusion $e^{-bD_0}$ is indicated by a dotted line, whereas the low-$b$, short-time approximation $e^{-bD}$, where $D$ is given by Eq. \eqref{eq:short}, is plotted as a dashed-dotted line. The high-gradient asymptotic regime \eqref{eq:localization_slab} appears as a dashed line.
}
\label{fig:slab}
\end{figure}

The signals are presented in Fig. \ref{fig:slab}. At low gradients, the Gaussian phase approximation is valid with an effective diffusion coefficient $D$. Note, however, the deviation from the free diffusion signal $E=e^{-bD_0}$ due to restriction by the slab. A short-time analysis \cite{Mitra1992a,Mitra1993a,Swiet1994a,Moutal2019a} yields an apparent diffusion coefficient
\begin{equation}
D\approx D_0\left(1-\eta\frac{4}{3\sqrt{\pi}}\frac{S}{V}\sqrt{D_0 T}\right)\;,
\label{eq:short}
\end{equation}
where $S/V$ is the surface-to-volume ratio of the confining domain ($S/V=2/L$ for a slab), $\eta\approx 0.9$ is a numerical prefactor that depends on the sequence, and $T=\Delta+\delta$ is the duration of the gradient sequence.

This is a short-time approximation in the sense that $\eta\sqrt{D_0 T}/L$ should be small enough. In addition, this formula relies on the Gaussian phase approximation which requires small $b$-values.
For our parameters, we get $D\approx 0.66 D_0$ and the agreement between $e^{-bD}$ and the signal at low gradients is good.

At higher gradients, the signal deviates significantly from the Gaussian regime and follows the asymptotic decay (see Eq. \eqref{eq:signal_localization} and Refs. \cite{Grebenkov2007a,Stoller1991a,Huerlimann1995a,Swiet1994a,Grebenkov2018a})
\begin{equation}
E_{\rm slab}\approx 2C_{1,1}\exp(-\lvert a'_1\rvert \ell_\delta^2/\ell_g^2)\;,
\label{eq:localization_slab}
\end{equation}
where the prefactor $C_{1,1}$ can be computed exactly \cite{Huerlimann1995a,Grebenkov2014b} and scales as $\ell_g/L$. One can interpret this prefactor as the fraction of spins inside the two layers of thickness $\sim \ell_g$ where the signal is localized. 
As we discussed in Sec. \ref{section:theory_diffusion}, $C_{1,1}$ decreases with increasing diffusion step duration $\Delta-\delta$.
Note that the localization regime emerges at gradients as small as $10~\rm{mT/m}$. Since the width of the slab is much greater than $\ell_g$, one can treat the localization layers on both sides of the slab as independent from each other (see Fig. \ref{fig:overlapping} and the discussion in Sec. \ref{section:theory_geometry}).

We observe a remarkable agreement between the experimental data, exact solution via the matrix formalism, and the asymptotic relation \eqref{eq:localization_slab}. Note that the latter contains no fitting parameter (the prefactor $C_{1,1}$ was found by computing the eigenmodes numerically). Systematic minor deviations of the experimental points may be caused by weak misalignment of the gradient direction (i.e. $\theta$ slightly different from $0$ in Eq. \eqref{eq:angle_slab}) or weak surface relaxivity. Note that we performed all computations with a square gradient profile instead of a trapezoidal one (in other words, with $\epsilon=0$, see Fig. \ref{fig:PGSE}) and we checked that not accounting for the trapezoidal profile had a negligible influence on the computed signal due to the very short ramp-up time ($\epsilon=0.34~\rm{ms}$).

\subsection{Diffusion inside a cylinder}

\begin{figure}[t]
\centering
\includegraphics[width=0.49\linewidth]{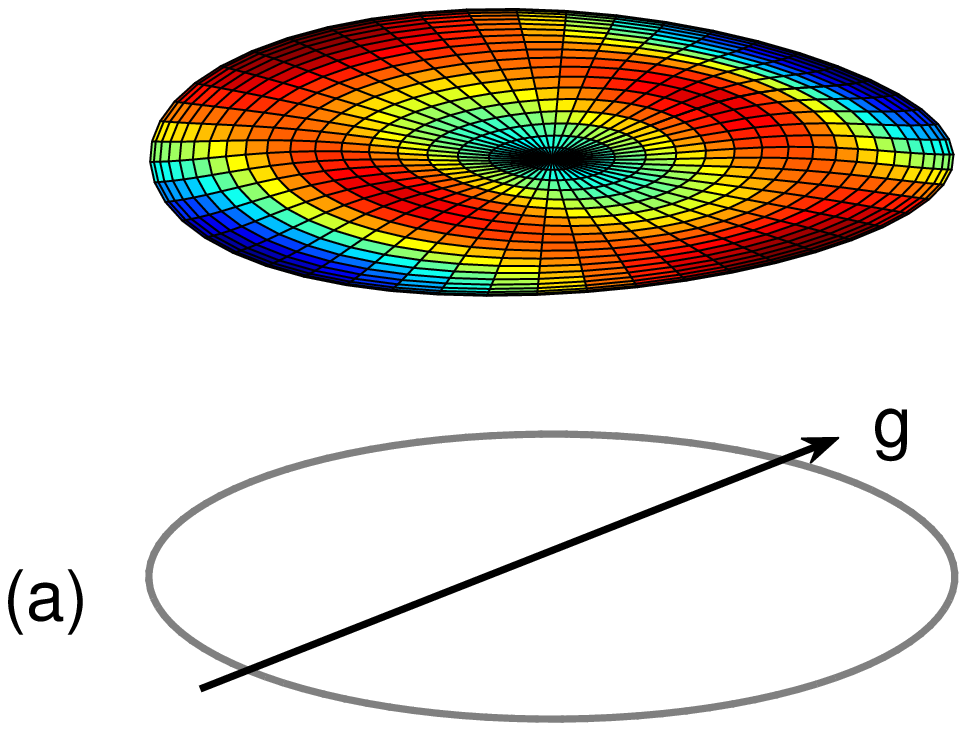}
\includegraphics[width=0.49\linewidth]{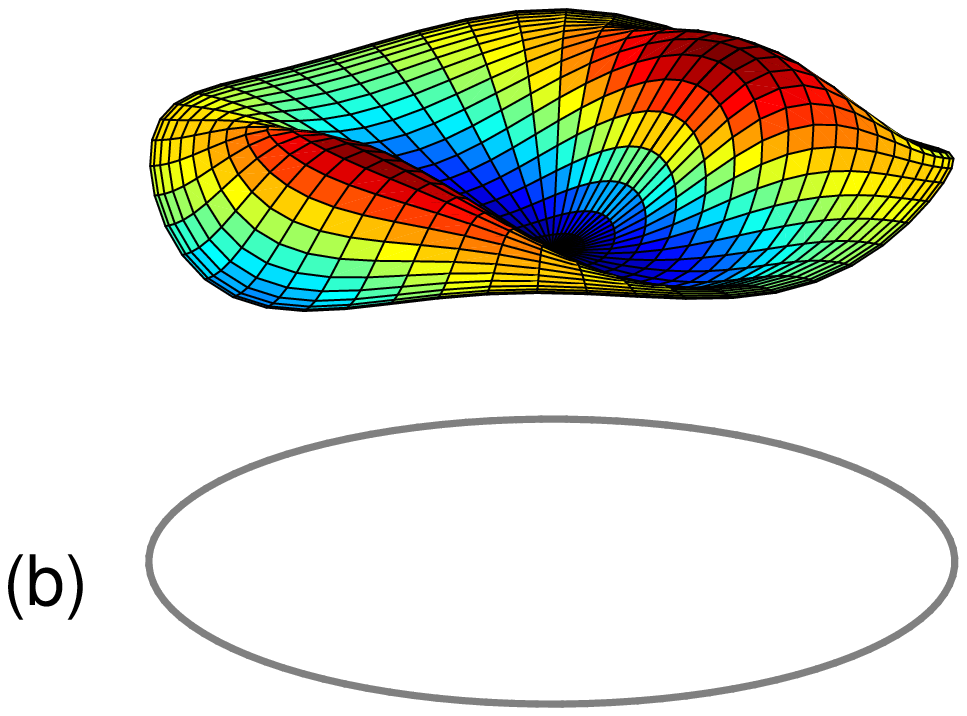}
\includegraphics[width=0.49\linewidth]{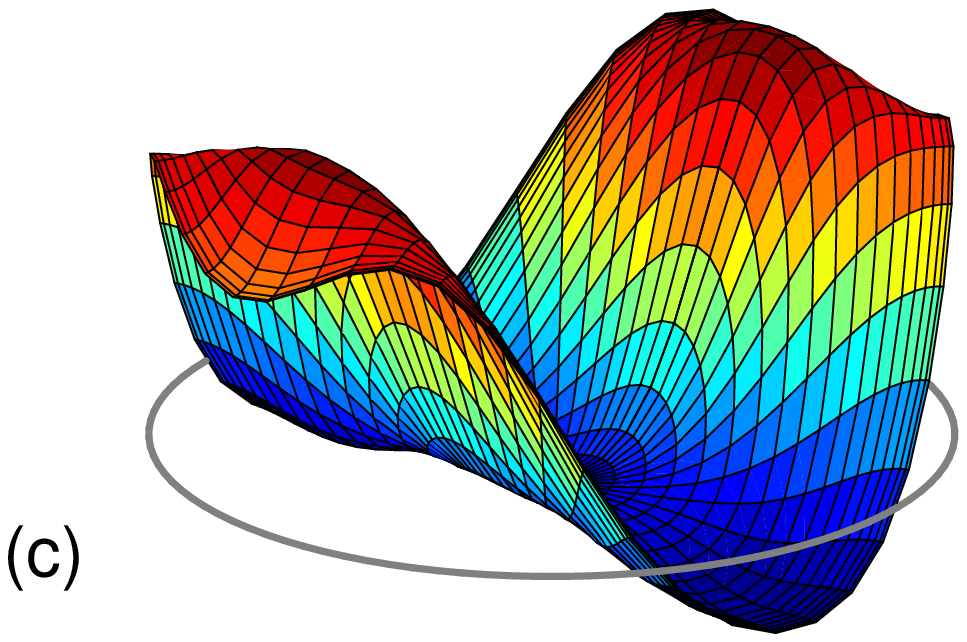}
\includegraphics[width=0.49\linewidth]{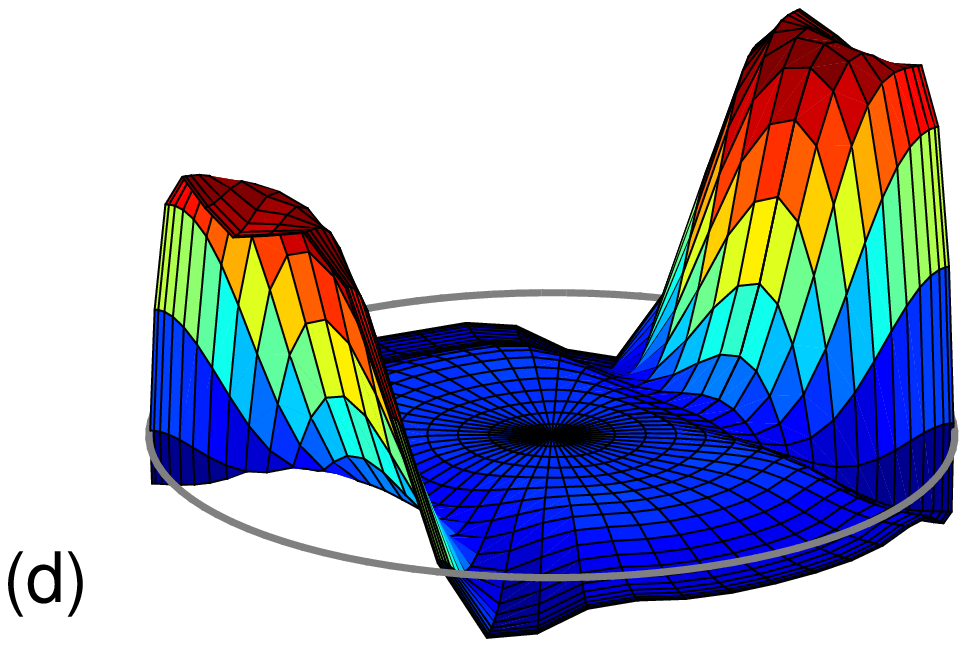}
\caption{Transverse magnetization computed by matrix formalism inside phantom $2$ (a cylinder of diameter $3.8~\rm{mm}$) for four values of the gradient $g$: (a) $2~\rm{mT/m}$, (b) $5~\rm{mT/m}$, (c) $10~\rm{mT/m}$ and (d) $32~\rm{mT/m}$. The direction of the gradient is indicated by an arrow.}
\label{fig:cylinders_magn}
\end{figure}

If the diameter of the cylinder is much larger than the gradient length $\ell_g$ and the diffusion length $\ell_\delta$, then the cylinder geometry is locally similar to the slab geometry except for the smaller ``localization pocket'' where the magnetization does not vanish. An argument similar to that of Eq. \eqref{eq:angle_slab} implies that the magnetization is localized where the angle between the gradient and the normal vector to the boundary is close to $0$ (this condition becomes more and more restrictive at higher $b$-values according to Eqs. \eqref{eq:angle_slab} and \eqref{eq:localization_slab}). The transverse magnetization inside a cylinder, obtained with a matrix formalism computation, supports this argument (Fig. \ref{fig:cylinders_magn}). In fact, as the gradient increases, the magnetization gradually transforms from a flat uniform profile to the one that is localized around two opposite points on the boundary of the cylinder and displays two independent pockets at sufficiently high gradients.

\begin{figure}[t]
\centering
\includegraphics[width=0.99\linewidth]{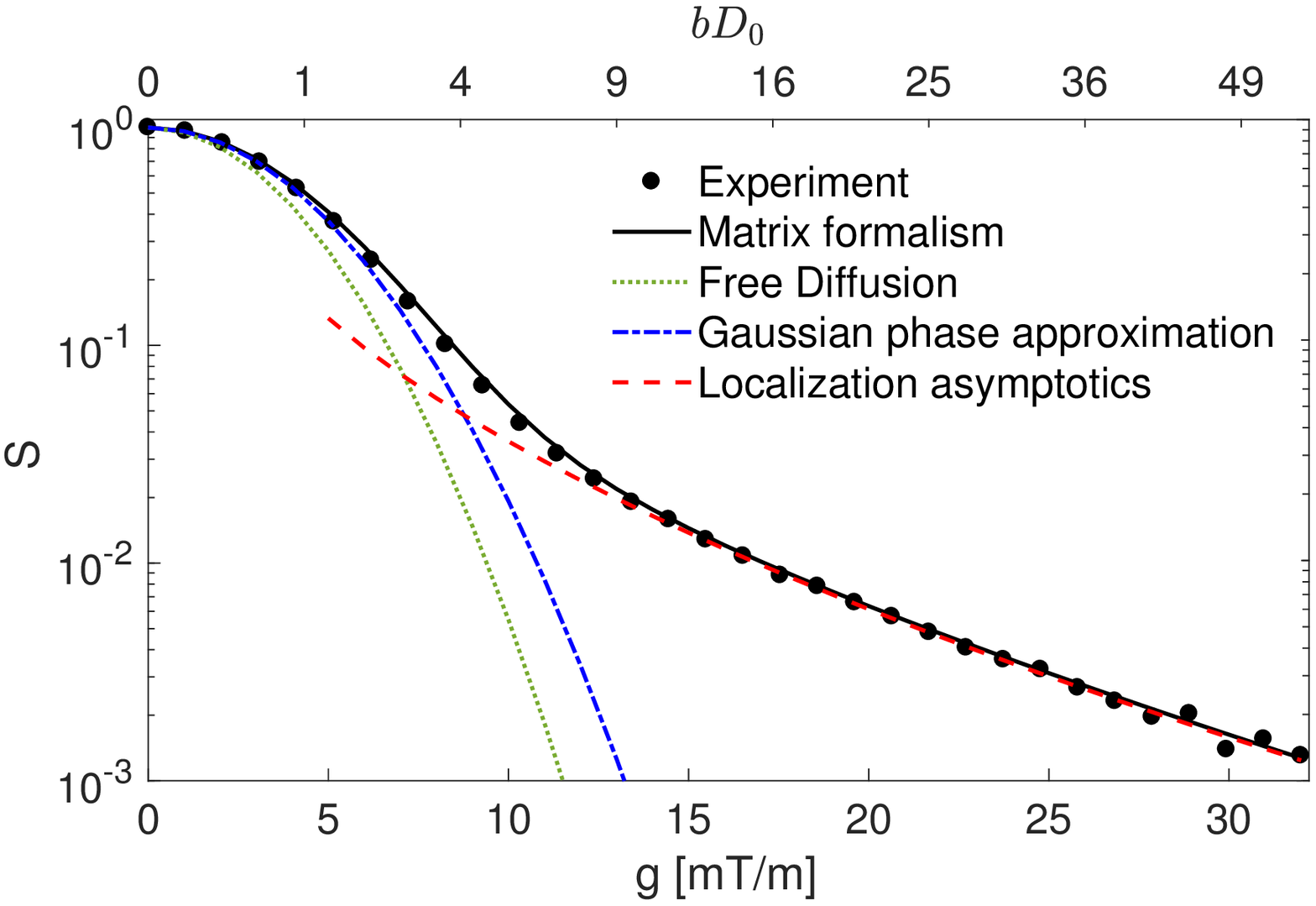}
\caption{Signal attenuation for diffusion inside phantom $2$ (a cylinder of diameter $3.8~\rm{mm}$). Experimental results are shown by full circles and matrix formalism computation by a solid line. The signal for free diffusion is indicated by a dotted line, whereas the short-time approximation $e^{-bD}$, where $D$ is given by Eq. \eqref{eq:short}, is plotted as a dashed-dotted line. The high-gradient asymptotic regime \eqref{eq:localization_cylinder} appears as a dashed line.}
\label{fig:cylinders}
\end{figure}

In Fig. \ref{fig:cylinders} we show the signal for diffusion inside a cylinder of diameter $3.8~\rm{mm}$. 
Although the signal decays faster than in the slab, one observes a similar stretched-exponential behavior. Using Eqs. \eqref{eq:lambda_1_next} and \eqref{eq:signal_localization}, one gets the asymptotic decay for the cylinder:
\begin{equation}
\!\!\!\!\!\!\! E_{\rm cyl}\approx C\exp\left(-\lvert a'_1\rvert \frac{\ell_\delta^2}{\ell_g^2} - \frac{\ell_\delta^2}{R^{1/2}\ell_g^{3/2}}+\frac{\sqrt{3}\ell_\delta^2}{2\lvert a'_1 \rvert R \ell_g}\right)\;,
\label{eq:localization_cylinder}
\end{equation}
where $C$ is given by Eq. \eqref{eq:overlapping}. Here, $R$ is large enough so that there is no overlapping between the first two eigenmodes, and $C=2C_{1,1}$. The prefactor $C_{1,1}$ can be computed numerically from the eigenmodes and scales approximately as $(\ell_g/R)^{3/2}$. One observes the perfect agreement between experiment, matrix formalism computation, and asymptotic formula at high gradients (without any fitting parameter).Note also that Eq. \eqref{eq:localization_slab} with only the leading term is not accurate here so that the correction terms in the exponential are indeed important.

\begin{figure}[t]
\centering
\includegraphics[width=0.99\linewidth]{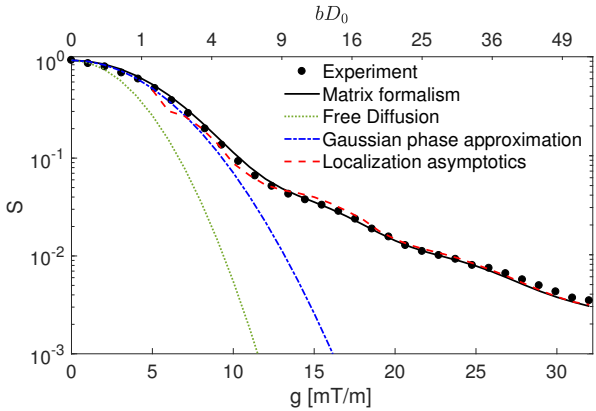}
\caption{Signal attenuation for diffusion inside phantom $3$ (a cylinder of diameter $2~\rm{mm}$). Experimental results are shown by full circles and matrix formalism computation by a solid line. The signal for free diffusion is indicated by a dotted line, whereas the short-time approximation $e^{-bD}$, where $D$ is given by Eq. \eqref{eq:short}, is plotted as a dashed-dotted line. The high-gradient asymptotic regime \eqref{eq:localization_cylinder} appears as a dashed line.}
\label{fig:cylinders2}
\end{figure}

\begin{figure}[t]
\centering
\includegraphics[width=0.49\linewidth]{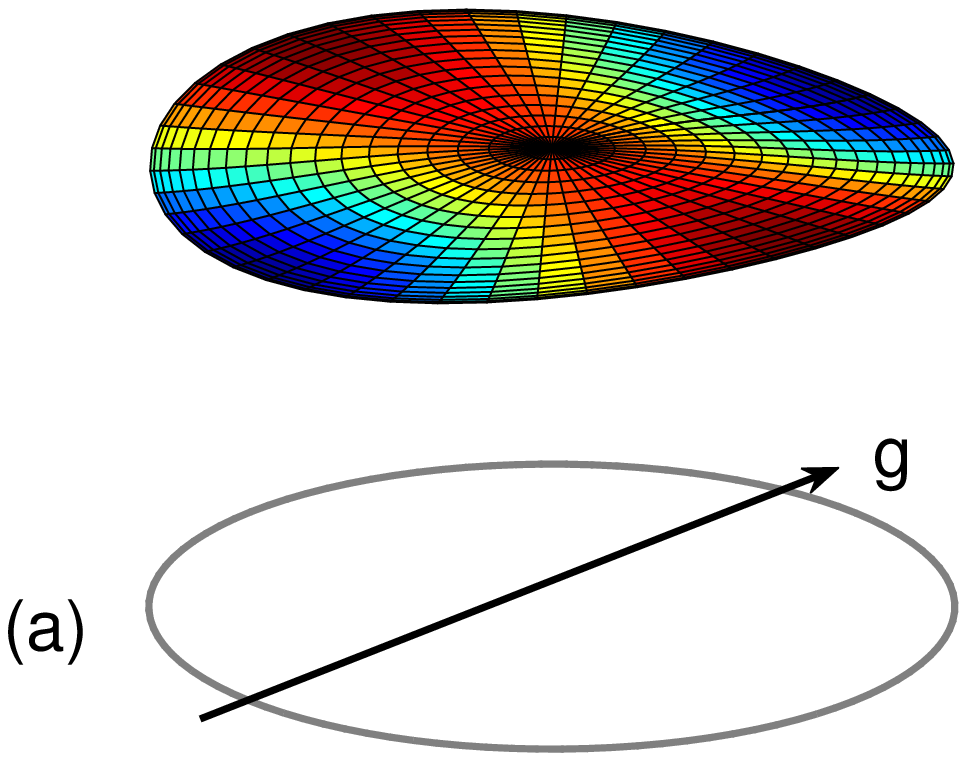}
\includegraphics[width=0.49\linewidth]{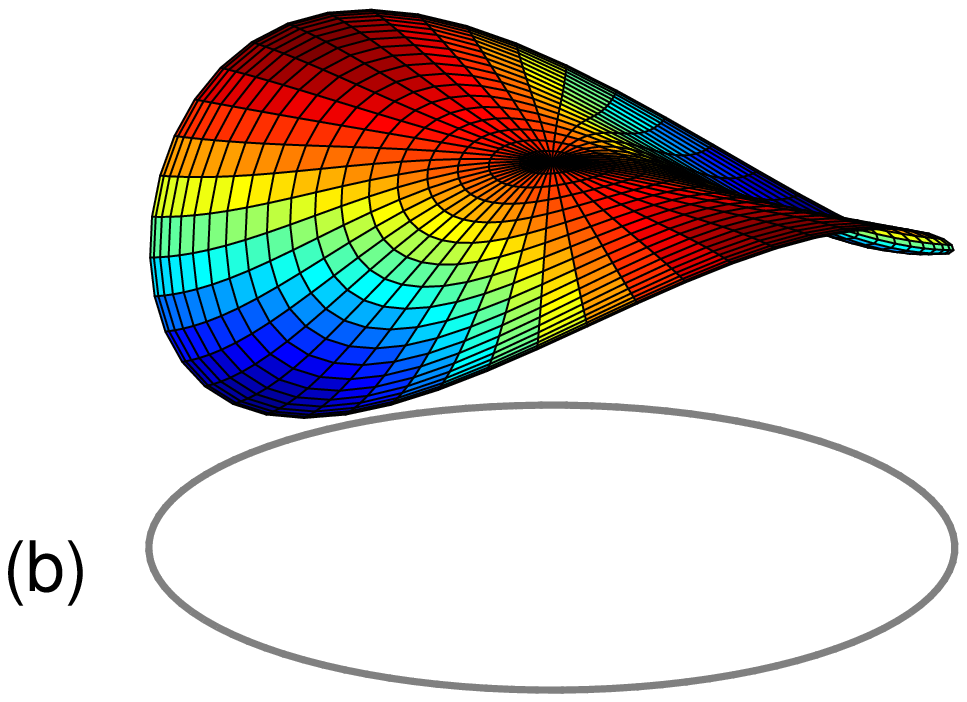}
\includegraphics[width=0.49\linewidth]{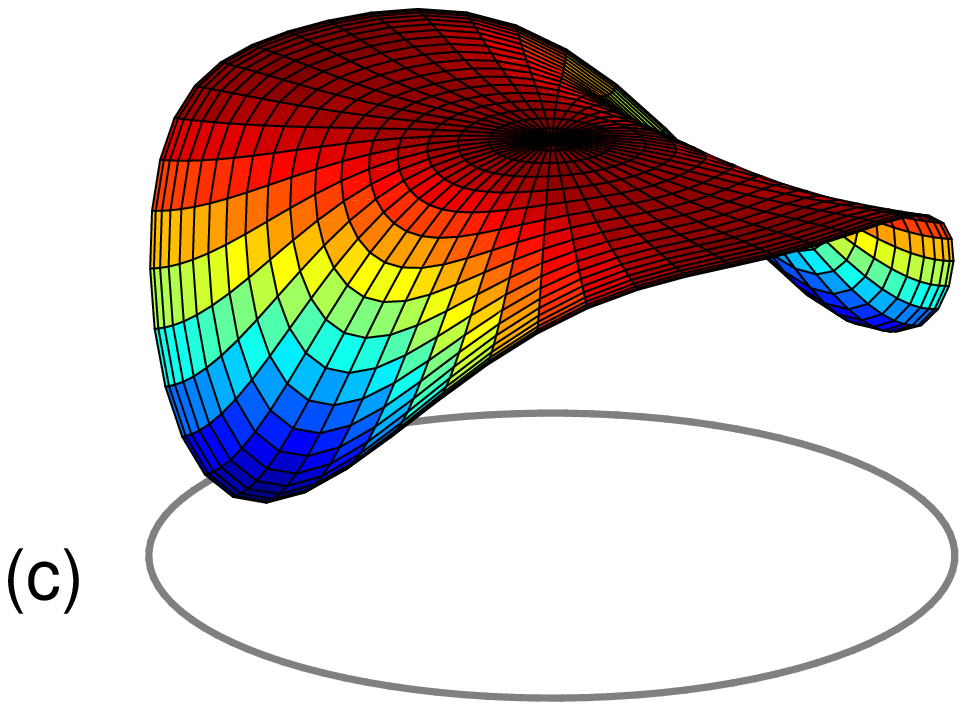}
\includegraphics[width=0.49\linewidth]{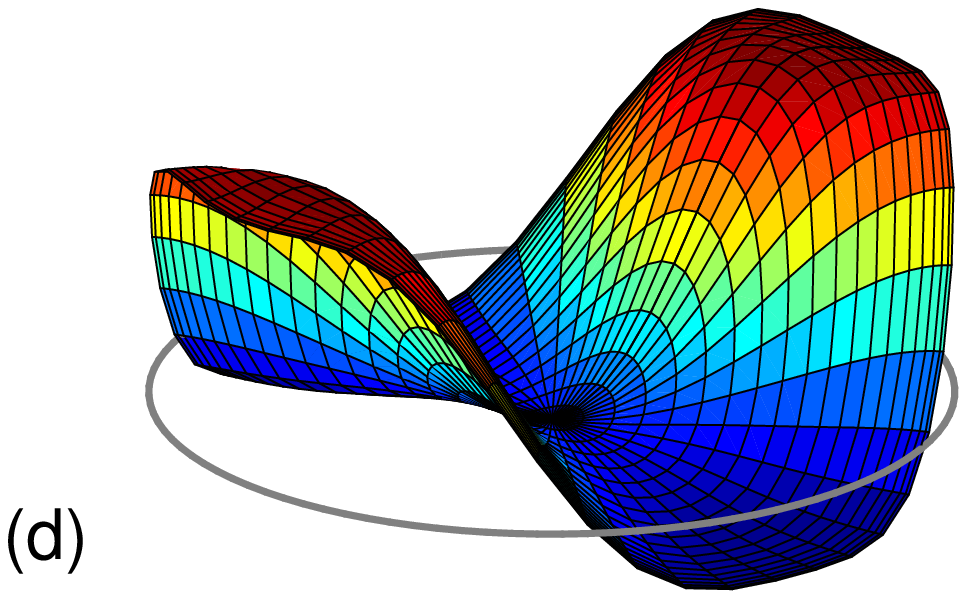}
\caption{Transverse magnetization computed by matrix formalism inside phantom $3$ (a cylinder of diameter $2~\rm{mm}$) for four values of the gradient $g$: (a) $2~\rm{mT/m}$, (b) $5~\rm{mT/m}$, (c) $10~\rm{mT/m}$ and (d) $32~\rm{mT/m}$. The direction of the gradient is indicated by an arrow.}
\label{fig:cylinders2_magn}
\end{figure}

For a cylinder of a smaller diameter ($2R=2~\rm{mm}$, see Fig. \ref{fig:cylinders2}), the signal shows some oscillations that are reminiscent of diffusion-diffraction patterns \cite{Callaghan1991b,Callaghan1992a,Callaghan1995a,Linse1995a,Callaghan1997a}. 
This is the case discussed in Sec. \ref{section:theory_geometry} where two localization pockets overlap because $\ell_g/(2R)$ is not small enough.
The signal is still given by Eq. \eqref{eq:localization_cylinder} but one cannot neglect the cross-term $C_{1,2}$ in the expression \eqref{eq:overlapping} of $C$.
The oscillations are then described by $C_{1,2}$ and appear on top of the asymptotic stretched-exponential decay.
These oscillations shown in Fig. \ref{fig:cylinders2} are very well reproduced by the asymptotic formulas \eqref{eq:overlapping_B} and \eqref{eq:lambda_im} with $L=2R$, where the coefficients $\mu_1$, $\mu_2$, and $\beta_{1,2}$ were computed numerically from the eigenmodes. We stress that these coefficients generally depend on $g$ and that $\beta_{1,2}$ additionally depends on the diffusion step duration $\Delta-\delta$.
This overlapping phenomenon is supported by Fig. \ref{fig:cylinders2_magn} which illustrates that the magnetization inside the cylinder is not well localized even at the highest gradient available.

\subsection{Diffusion outside an array of rods}


The geometry of phantoms $4$ and $5$ is defined by $L$, the center-to-center spacing between rods, and $2R$, the diameter of the rods. We consider three different cases: (a) phantom $4$ ($L=4~\rm{mm}$ and $2R=3.2~\rm{mm}$) with the gradient vector in the diagonal direction; (b) phantom $5$ ($L=3.4~\rm{mm}$ and $2R=3.2~\rm{mm}$) and gradient vector in the diagonal direction; (c) phantom $5$ and gradient vector in the horizontal direction.

\begin{figure}[t]
\centering
\includegraphics[width=0.6\linewidth]{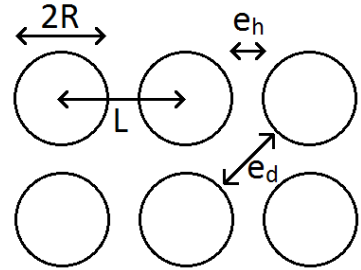}
\caption{Schematic representation of the rods showing the lengths $e_{\rm d}$ and $e_{\rm h}$.}
\label{fig:schema}
\end{figure}

\begin{figure}[hpbt]
\centering
\includegraphics[width=0.98\linewidth]{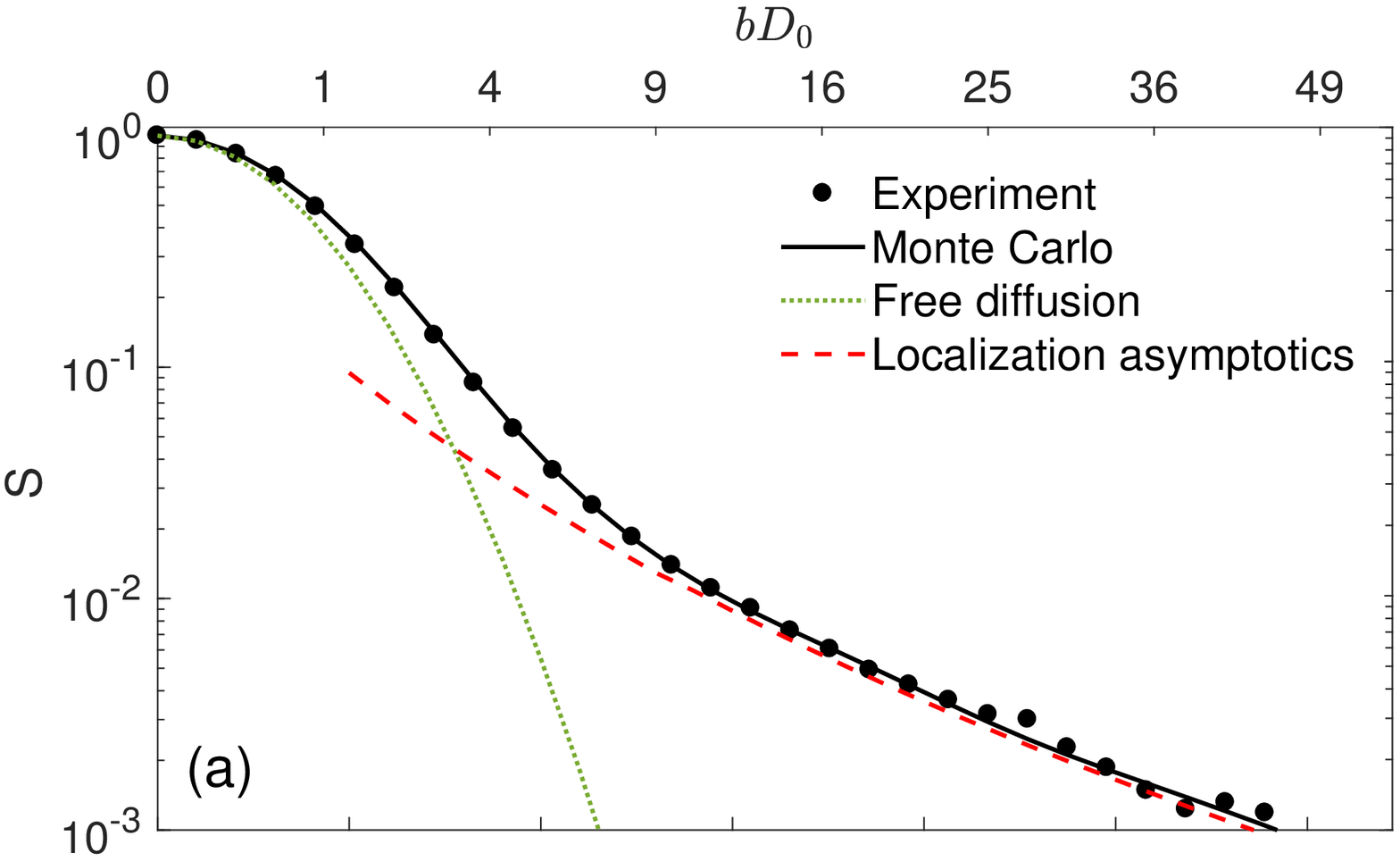}
\includegraphics[width=0.98\linewidth]{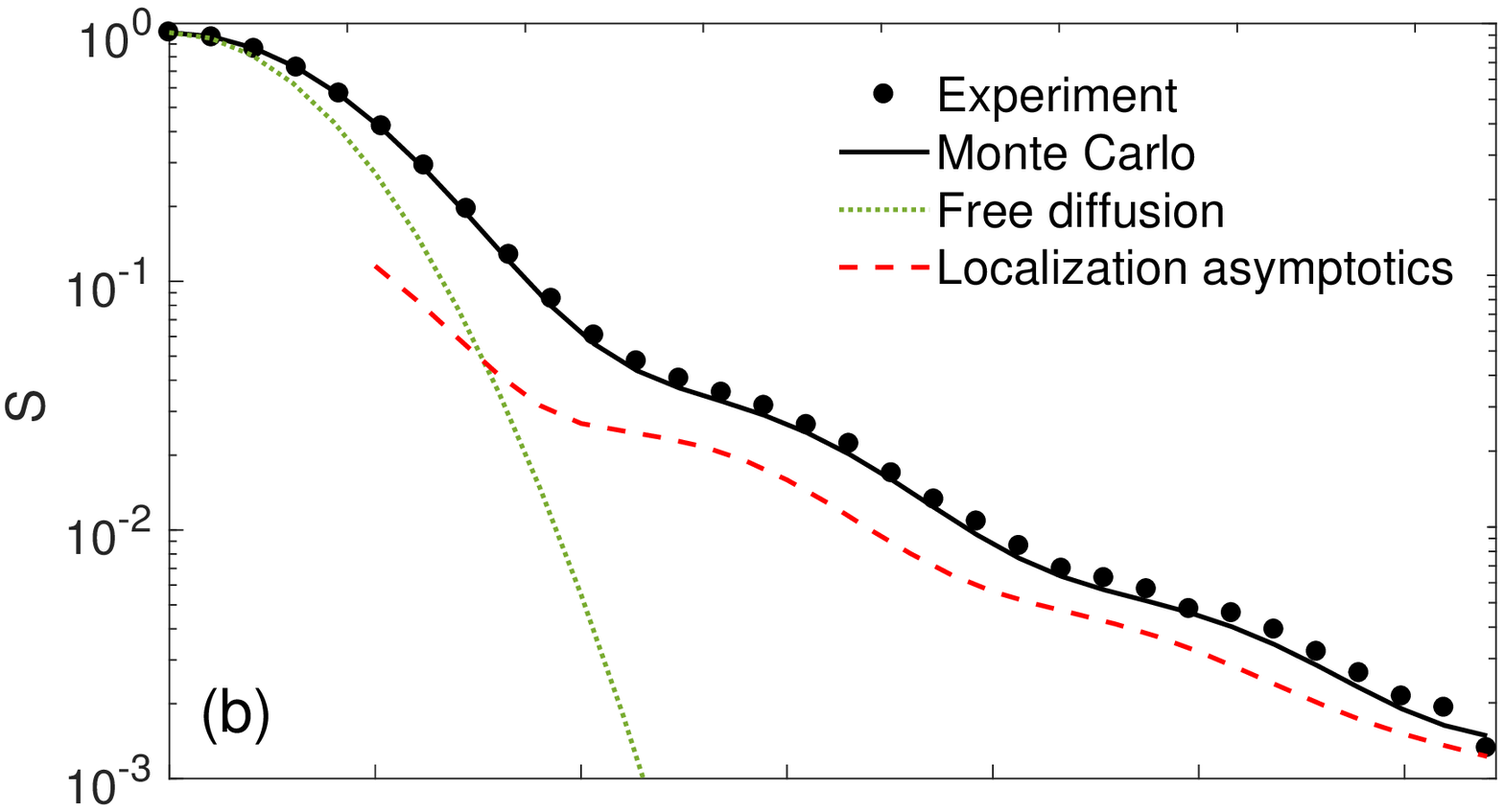}
\includegraphics[width=0.98\linewidth]{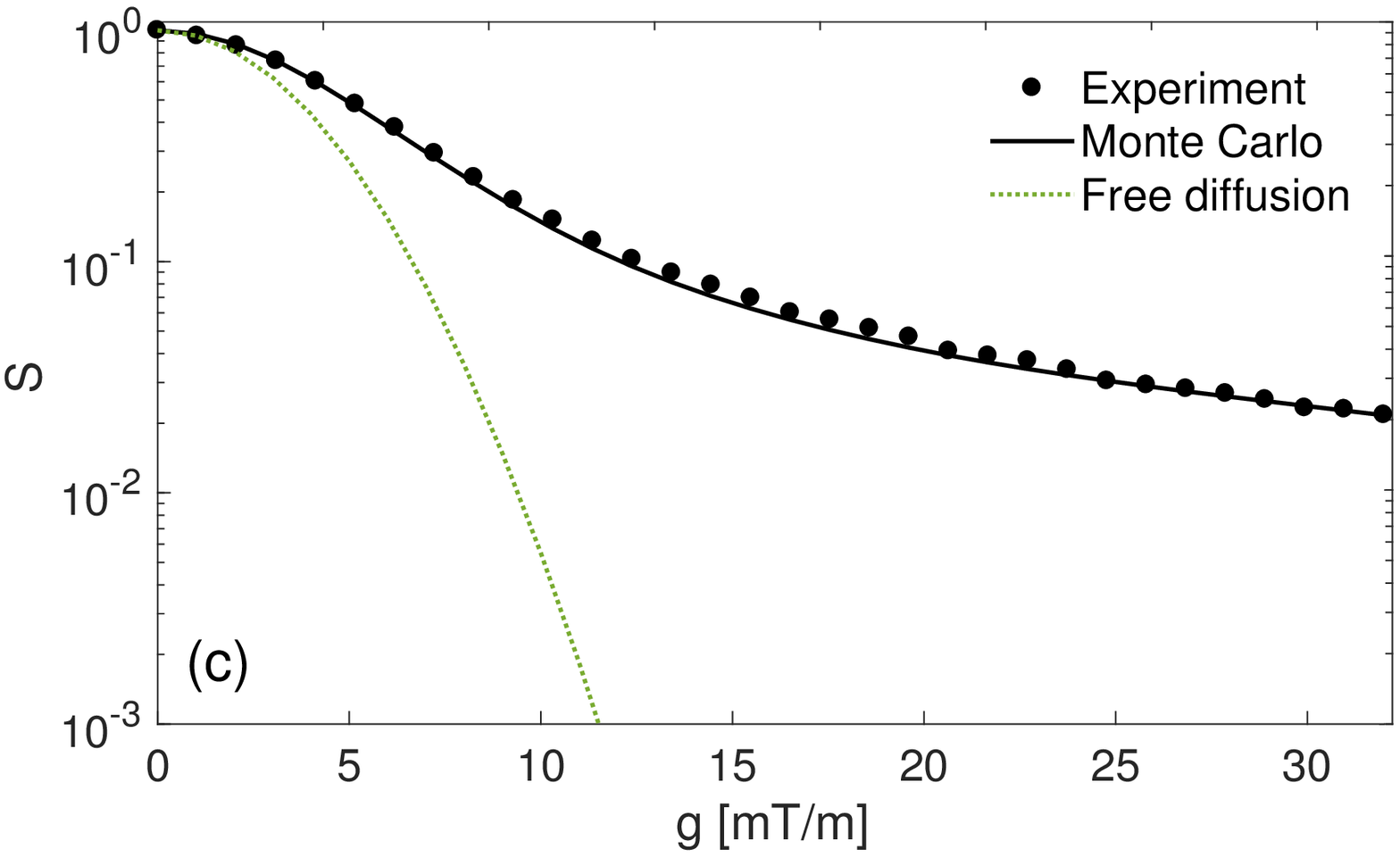}
\caption{Signal attenuation for diffusion in phantoms $4$ and $5$ (array of rods of diameter $3.2~\rm{mm}$ and center-to-center spacing $4~\rm{mm}$ and $3.4~\rm{mm}$, respectively).
(a) phantom $4$ with the gradient in the diagonal direction, spacing $e_{\rm pock}=2.5~\rm{mm}$; (b) phantom $5$ and diagonal gradient direction, spacing $e_{\rm pock}=1.6~\rm{mm}$; (c) phantom $5$ and horizontal gradient direction, spacing $e_{\rm pock}=0.2~\rm{mm}$.
Experimental results are shown by full circles and Monte Carlo simulations by a solid line. The signal for free diffusion is indicated by a dotted line. The asymptotic regime \eqref{eq:localization_rods} appears as a dashed line. It is not shown for (c) as this regime is not applicable here, see the text.}
\label{fig:rods}
\end{figure}

The main difference between these three cases is the spacing $e_{\rm pock}$ between two neighboring rods along the gradient direction, i.e. the spacing between two neighboring localization pockets (see Fig. \ref{fig:schema}): $e_{\rm pock}=e_{\rm d}=\sqrt{2}L-2R=2.5~\rm{mm}$ in (a); $e_{\rm pock}=e_{\rm d}=1.6~\rm{mm}$ in (b); and $e_{\rm pock}=e_{\rm h}=0.2~\rm{mm}$ in (c). Figure \ref{fig:rods} shows the signal for these three cases, ordered by descending $e_{\rm pock}$. Note that here the signal is formed by the magnetization localized near the rods and by the magnetization localized near the borders of the casing in which the phantom is enclosed.
We did not plot the low-$b$, short time approximation $E=e^{-bD}$ here because the surface-to-volume ratio of the structure is too large so that the approximate formula \eqref{eq:short} for $D$ is not valid.

First of all, one can note an excellent agreement between experimental data and Monte Carlo simulations.
The high-$g$ asymptotic behavior of the eigenvalues and the signal in a rods geometry is similar to the one for cylinders in Eq. \eqref{eq:localization_cylinder} except for a sign change due to the opposite curvature \cite{Grebenkov2018b}:
\begin{equation}
\!\!\!\!\!\!\!E_{\rm rods}\approx C\exp\left(-\lvert a'_1\rvert \frac{\ell_\delta^2}{\ell_g^2} - \frac{\ell_\delta^2}{R^{1/2}\ell_g^{3/2}}-\frac{\sqrt{3}\ell_\delta^2}{2\lvert a'_1 \rvert R \ell_g}\right)\;,
\label{eq:localization_rods}
\end{equation}
where $C$ is given by Eq. \eqref{eq:overlapping} and may be computed numerically from the eigenmodes. This formula matches very well the signal at high gradients in case (a) (see Fig. \ref{fig:rods} (a)).

In the previous subsection, we already saw the signal without oscillations (Fig. \ref{fig:cylinders}) due to well-localized states, as well as the signal with oscillations (Fig. \ref{fig:cylinders2}) due to a partial overlap of two localization pockets when the size of the confining domain is not very large compared to $\ell_\delta$ and $\ell_g$. Here the same phenomenon occurs.
With $e_{\rm pock}$ much larger than $\ell_g$ (case (a)), there is little overlapping between the localization pockets of neighboring rods. This ensures the localization of the eigenmodes and the small amplitude of the oscillations in the signal (see Sec. \ref{section:theory_geometry}), i.e. the validity of $C=2C_{1,1}$.
In the case where the ratio between $e_{\rm pock}$ and $\ell_g$ is smaller, i.e. localization pockets overlap more (case (b)), more pronounced oscillations on top of the overall decay \eqref{eq:localization_rods} arise (see Fig. \ref{fig:rods} (b)). 
The signal is still described by Eq. \eqref{eq:localization_rods}, and the oscillations are contained in the cross-term $C_{1,2}$ from the expression \eqref{eq:overlapping} of $C$ and may be computed from Eqs. \eqref{eq:overlapping_B} and \eqref{eq:lambda_im} (with $L=e_{\rm pock}$).
Systematic deviations between the exact signal and the asymptotic formulas may be attributed to the truncation of Eq. \eqref{eq:lambda_1_next}, neglecting higher-order modes in the expression of the signal, and not accounting for the borders in the experimental setup.

In turn, in case (c) $e_{\rm pock}$ is smaller than $\ell_g$ even at the highest gradient available. If one could approximate the small space between two neighboring rods as a slab, then the ratio $\ell_g/e_{\rm pock}$ would be too large for localization to be relevant (see Sec. \ref{section:theory_geometry}). Our conjecture is that the residual signal at high gradients may be interpreted as a kind of motional narrowing inside the small gaps of width $e_{\rm pock}$ between the rods (see \cite{Robertson1966a,Grebenkov2007a}).

\section{Discussion and Conclusion}
\label{section:conclusion}

We have observed and described the localization regime in three geometries: slab, cylinder, and array of circular obstacles (rods). The localization regime appears whenever the gradient length $\ell_g=(\gamma g/D_0)^{-1/3}$ is much smaller than the diffusion length $\ell_\delta=(D_0\delta)^{1/2}$ and any relevant geometrical length scale $\ell_s$ of the medium along the gradient direction. Thus, it is universal at high gradients and non-narrow pulses. In this regime, the transverse magnetization is localized near the obstacles, boundaries, or membranes of the sample. For this reason, the signal is particularly sensitive to the microstructure of the medium. In particular, possible oscillations of the signal are caused by a partial overlap between localized magnetization pockets and thus contain information about mutual arrangements of obstacles.

Let us clarify the role of the conditions (i) $\ell_g \ll \ell_\delta$ and (ii) $\ell_g\ll\ell_s$. Condition (i) ensures that the eigenmode decomposition of the transverse magnetization may be truncated to its first terms. In turn, the signal decays exponentially with $\delta$ and its dependence on $g$ is essentially determined by the first eigenvalue of the Bloch-Torrey operator.
Furthermore, condition (ii) is necessary for the localization of the eigenmodes of the Bloch-Torrey operator and the validity of the high-$g$ expansion of its eigenvalues. Thus, both conditions are required for the localization of the transverse magnetization and the stretched-exponential decay of the signal with $g$.

An extreme case where condition (i) is not satisfied would be narrow gradient pulses experiments ($g\to \infty$ and $\delta \to 0$). Although these experiments require high gradients, they do not achieve the localization regime. In fact, whereas the localization regime emerges when $(\gamma g)^2 \delta^3 D_0 \gg 1$, narrow-pulse experiments correspond to $\gamma g \to \infty$ and $\delta \to 0$ such that $\gamma g \delta=q$ is a finite value. This leads to $(\gamma g)^2 \delta^3 D_0 \to 0$. In other words, the signal attenuation is not produced by the encoding step but by the subsequent diffusion step with $g=0$.
This is evident from the fact that the signal is left unchanged if one sets $\Delta=0$, i.e. no diffusion time between two short gradient pulses.
On the other hand, condition (ii) is typically not satisfied in the motional narrowing regime ($g \to 0$ and $\delta \to \infty$). Then, the eigenmodes of the Bloch-Torrey operator are close to the eigenmodes of the Laplace operator. In particular, the first Laplacian eigenmode is constant if one assumes impermeable, non-relaxing boundaries. Therefore, the transverse magnetization at long times is uniform inside the sample.


Slab and cylinder are confined geometries that may model an intracellular space. It is well-known that such domains produce non-Gaussian signals, for example in the limit of narrow pulses (e.g. diffusion-diffraction patterns, see \cite{Callaghan1991b,Callaghan1992a,Callaghan1995a,Linse1995a,Callaghan1997a}). However, the signal from extracellular space is almost always assumed to be Gaussian, and non-Gaussian effects are attributed to multiple contributing pools. In other words, one assumes that the obstacles may be treated as an effective medium with an effective diffusivity $D$. The measurements performed on the array of rods show that the extracellular signal is not Gaussian at high gradients and that it simply results from the localization of the signal at the outer boundaries of the obstacles. Ignoring this effect may lead to false interpretations created by commonly used fitting models.

We stress that the localization regime in our setting starts to emerge with moderate gradients of about $10~\rm{mT/m}$, or at $bD_0$ about $4$. These gradients are easily achieved in most clinical scanners. Note that the localization regime emerges under the condition $\ell_\delta/\ell_g \gg 1$ which can be restated as $bD_0 \propto \gamma^2 D_0 g^2 \delta^3 \gg 1$. In order to rescale the experimental conditions from xenon gas to water, we compute the ratio $(\gamma^2 D_0)_{\rm xenon}/(\gamma^2 D_0)_{\rm water} \approx 10^3$. This means that in order to have the same value of $\gamma^2 D_0 g^2 \delta^3$, one has to increase $g$ and $\delta$ such that $g^2 \delta^3$ is $10^3$ times larger with water than with xenon. For example, the experiments by H{\"u}rlimann \textit{et al} were performed with $\delta=60~\rm{ms}$ which is approximately $10$ times longer than in our experiments, and gradients of comparable magnitude as ours (around $20~\rm{mT/m}$).

Theoretical aspects of diffusion NMR have been broadly explored at weak gradients by means of perturbation theory. While surface-to-volume ratio,  overall curvature, permeability and surface relaxivity could be estimated  at short times \cite{Mitra1992a,Mitra1993a,Moutal2019a}, the structural organization of barriers is accessible at long times \cite{Novikov2011a,Novikov2014a}. In biomedical applications, the apparent diffusion coefficient and kurtosis are employed as biomarkers to detect stroke, tumors, lesions, partial tissue destruction, etc. (see \cite{Novikov2018b} and references therein). In spite of its considerable progress in medicine and material sciences, the comprehensive theory of diffusion NMR remains to be elaborated. The recent mathematical advances show that the gradient term presents a singular perturbation to the Laplace operator in unbounded domains, resulting in the discrete spectrum of the Bloch-Torrey operator \cite{Almog2018a}. But even in bounded domains, the presence of the gradient term may lead to branching points of the spectrum and thus a finite radius of convergence of the cumulant expansion \cite{Stoller1991a}. These mathematical evidences urge for revising the conventional perturbation theory paradigm and developing non-perturbative approaches to diffusion NMR. Even so the localization regime is yet poorly understood and exploiting its potential advantages is still challenging in experiments (partly due to strongly attenuated signals), the high sensitivity of the signal to the microstructure at high gradients is a promising avenue for creating new experimental protocols. If former theoretical efforts were essentially focused on eliminating the dephasing effects and reducing the mathematical problem to pure diffusion, future developments have to aim at exploiting the advantages of high gradients and accentuating the differences from pure diffusion.

\end{document}